\documentclass[superscriptaddress,groupedaddress,nofootnoteinbib,11pt]{article}
\pdfoutput=1
\usepackage{color}
\usepackage{graphicx}
\usepackage{dcolumn}
\usepackage{bm}
\usepackage{amssymb}
\usepackage{amsmath} 
\usepackage{sectsty}
\usepackage{colortbl}
\usepackage{latexsym}
\usepackage{float}
\usepackage{ifthen}
\usepackage{enumerate}
\usepackage{url}
\usepackage[force]{feynmp-auto}
\usepackage{jcappub}
    \usepackage{picinpar}
    \usepackage{colortbl}
\usepackage{multirow}
	\usepackage{float}
	      \usepackage{setspace}
\usepackage{array}
\usepackage{bm}
\usepackage{amsopn}
\usepackage{tablefootnote}
\renewcommand{\vec}[1]{\bm{\mathrm{{#1}}}}
\usepackage{booktabs}
\usepackage[table]{xcolor}

\usepackage[utf8]{inputenc}
\usepackage{array}
\usepackage{makecell}

\definecolor{lightgray}{gray}{0.9}

\newcommand{\p}{\partial}

\def\ba{\begin{eqnarray}}
\def\ea{\end{eqnarray}}
\def\beq{\begin{eqnarray}}
\def\eeq{\end{eqnarray}}
\def\noi{\noindent}

\def\({\left(}
\def\){\right)}

\def\p{\partial}

\def\<{\langle}
\def\>{\rangle}

\definecolor{verde}{rgb}{0,0.5,0}
\def\blu{\color{blue}}

\usepackage{titlesec}

 \def\bea  {\begin{eqnarray}}   \def\eea  {\end{eqnarray}}

\newboolean{editorial}
\setboolean{editorial}{true}
\newcommand{\editorial}[2]{\ifthenelse{\boolean{editorial}}

{\textcolor{red}{ [\textsf{\textbf{{#1}}}:} \textcolor{blue}{\textsf{{#2}}}\textcolor{blue}{]}}{}}

\definecolor{dullpurple}{rgb}{0.431,0.188,0.534}
\definecolor{darkgreen}{rgb}{0.133,0.545,0.133}
\definecolor{verde}{rgb}{0,0.5,0}

\newboolean{comment}
\setboolean{comment}{true}
\newcommand{\comment}[2]{\ifthenelse{\boolean{comment}}{\textcolor{blue}{{{{#1}}}: }\textcolor{verde}{{#2}}}{}\\}

\newboolean{red2}
\setboolean{red2}{true}
\newcommand{\redtwo}[2]{\ifthenelse{\boolean{red2}}{\textcolor{red}{#1}}\\}

\begin{document}

\title{
Tensor non-Gaussianities from Non-minimal Coupling to the Inflaton}

\author{Emanuela Dimastrogiovanni$^{a}$,  Matteo Fasiello$^{b}$, Gianmassimo Tasinato$^{c}$, and David Wands$^{b}$}
\affiliation{$^{a}$ CERCA/Department of Physics, Case Western Reserve University, Cleveland, OH 44106, U.S.A.}
\affiliation{$^{b}$ Institute of Cosmology and Gravitation, University of Portsmouth, Dennis Sciama Building, Burnaby Road, PO1 3FX, U.K.}
\affiliation{$^{c}$ Department of Physics, Swansea University, Swansea, SA2 8PP, U.K.}

\abstract{Tensor non-Gaussianity represents an important future probe of the physics of inflation. Inspired by recent works, we elaborate further on the possibility of significant primordial tensor non-Gaussianities sourced by extra fields during inflation. Unitarity constraints  limit the impact of extra (spinning) particle content by means of a lower bound on the corresponding mass spectrum. For spin-2 particles, this takes the form of the well-known Higuchi bound.  Massive ($m\gtrsim H$) particles will typically decay during inflation unless they are non-minimally coupled to the inflaton sector: the inflating field  ``lifts" the dynamics of the extra field(s), effectively getting around the limits imposed by unitarity.  There exist several models that realize such a mechanism, but we focus here on the set-up of \cite{Bordin:2018pca} where, through an EFT approach, one is able to capture the essential features common to an entire class of theories. In the presence of an extra massive spin-2 particle, the interactions in the tensor sector mimic very closely those in the scalar sector of \textit{quasi-single-field} inflationary models. We calculate the tensor bispectrum in different configurations and extract its dependence on the extra tensor sound speed. We show in detail how one may obtain significant tensor non-Gaussianities whose shape-function interpolates between local and equilateral, depending on the mass of the extra field. We also estimate  the LISA response functions to a tensor bispectrum supporting  the intermediate-type shapes we find.
}

\maketitle

	\section{Introduction}
	\label{sec:introduction}

The existence of primordial gravitational waves is one of the key predictions of the inflationary paradigm \cite{inflation,reviewspgw}. In the simplest
  scenarios inflation is driven by a single scalar field with canonical kinetic terms, and the corresponding predictions on the tensor power spectrum are that of a small almost scale-invariant signal, slightly decreasing towards smaller length scales. In these set-ups vacuum fluctuations are the leading source of the signal and  the value of the Hubble parameter $H$ during inflation is the only unknown quantity: a detection would immediately reveal the energy scale of inflation. However, a growing literature \cite{growingliteratureaxions,growingliteraturescalars} of 
inflationary models points to the fact that predictions for primordial gravitational waves (PGW) can deviate dramatically from this simple picture. The crucial common ingredient is the presence of (a remnant of) extra degrees of freedom sourcing PGW.\\ Adding particle content to the inflationary Lagrangian can: 
\begin{itemize}
\item
break the one-to-one correspondence between $H$ and the energy scale of inflation; 
\item
greatly enhance the amplitude of the tensor signal; 
\item
violate consistency relations related to the squeezed
  limit of the tensor bispectrum; 
\item
  completely change the scale dependence of the power spectrum in favour of, for example, a blue spectrum or a 
  localised (sometimes termed ``bump")
  feature. 
   \end{itemize}
  The latter effect is particularly relevant in light of the possibility of detection at small scales via ground-based and space-based interferometers \cite{interferometers}. 
 Indeed, whilst assuming a constant tilt of the primordial tensor power spectrum one may extrapolate from the ESA Planck mission results \cite{Akrami:2018odb} (combined with LIGO and Virgo \cite{TheLIGOScientific:2016dpb}) tight constraints on small scales, there is much more scope to accommodate and test a signal exhibiting features across different scales. \\
Of course one ought to strive to ask not just ``what is possible" but also ``what is natural or compelling" according to some agreed upon criterion. Let us briefly mention one example. Axions and axion-like particles are ubiquitous in physics, from QCD \cite{Peccei:1977ur} to inflationary models \cite{Pajer:2013fsa}. In the latter context, the (approximate) shift symmetry of the axion potential is perhaps the most appealing property of such models: it makes it possible to tackle the so-called $\eta$-problem, i.e. the symmetry protects the inflaton mass under large quantum corrections \cite{Freese:1990rb}. Intriguingly, it tuns out that coupling an inflating axion to gauge fields during inflation is a way to make \textit{natural inflation} \cite{Freese:1990rb} more natural \cite{Pajer:2013fsa}. These extra gauge fields are a dissipation channel for the (originally too steep) inflaton potential. Remarkably, they may also become the leading source for PGW, delivering a chiral gravitational wave spectrum with a bumpy profile at certain scales \cite{bumpy} and potentially large tensor non-Gaussianities peaking in the equilateral configuration. We stress that this class of models is by no means an exhaustive or sufficiently representative set of the many more existing mechanisms for PGW generation -- see for example the recent \cite{Mylova:2018yap} for a new mechanism for enhancing
tensor modes at small scales,  exploiting  a non-attractor phase for single-field inflation. In general,
 we refer the interested reader to the review work in \cite{Guzzetti:2016mkm} for a more thorough survey of the literature.\\
In this work we shall embrace a rather different perspective on PGW generation. Instead of committing to a specific model, making a case for its naturalness in the model-building sense, and delivering its prediction, we will follow the general effective theory approach of \cite{Bordin:2018pca}. Much in the spirit of earlier work \cite{Cheung:2007st}, where an effective field theory of fluctuations around an inflating background was built for the single field case, the study in \cite{Bordin:2018pca} extends this analysis to include extra fields with sizable coupling to the inflaton sector. The non-trivial nature of the coupling is of particular relevance given that we are after enhancing the PGW spectrum.\\ Extra scalars, vectors, tensors and higher-spin degrees of freedom have all been enlisted to generate gravitational waves. Unlike the case of scalars and (Abelian) vector fields, for which the sourcing typically occurs at non-linear order, starting with tensors one may feed the gravitational wave signal already at first order. The trouble is that, as one goes up the spin ladder, unitarity is increasingly demanding on the mass spectrum of the particle content \cite{Higuchi:1986py,Fasiello:2013woa,Lee:2016vti}, for which constraints of the form $m\gtrsim H$ are typically in place.\\ Quantum fluctuations associated to a particle in that mass range will inevitably decay during inflation unless a non-minimal coupling to the inflaton sector sustains them. The effective Lagrangian put forward in \cite{Bordin:2018pca} includes the most general form  a similar non-trivial coupling might take\footnote{Under the assumption that the inflaton enjoys an approximate shift-symmetry.}. In what follows, we will rely on the interactions, starting already at quadratic order,  between the standard tensor sector and that of an additional spin-2 particle $\sigma$. Intriguingly, these are reminiscent of the dynamics characterising the scalar sector of quasi-single field inflation (QsF) \cite{Chen:2009zp}. Indeed, we recover many of the QsF features. The most relevant difference lies in the speed of propagation for the extra tensor (scalar) sector $\sigma$:  bounds on tensor (scalar) non-Gaussianity, do (not) allow a small sound speed. \\ 
\noi In this work we focus on the effects of the extra field content on the tensor sector\footnote{See also \cite{Biagetti:2017viz} and references therein for related models where the extra spin-2 field is included by means of the so-called \textit{dRGT} interactions.}. In particular, we examine tensor non-Gaussianities, an observable which can provide   a useful probe of the inflationary particle zoo for a wide range of scales, ranging from the CMB, to large-scale structure, all the way to interferometers. We show that couplings with
 additional tensor fields can   enhance  the amplitude of tensor non-Gaussianities, and lead to new shapes
 for the tensor bispectrum.\\ 

\noi The manuscript is organised as follows: in \textit{Section} \ref{sec:review} we briefly review the EFT approach that provides the quadratic and cubic interactions at the basis of our bispectrum calculation; in \textit{Section} \ref{sec:3} we show in detail the effect of an additional spin-2 particle on tensor non-Gaussianity, we derive the corresponding shape, its dependence on the mass of the extra field as well as its sound speed; in \textit{Section} \ref{sec:3} we also undertake the first step towards producing forecasts for the capability of LISA to detect tensor non-Gaussianities in models such as the one under study, we do so by computing the response function for bispectrum shapes that are intermediate between the local and the equilateral; in \textit{Section} \ref{sec:4} we discuss our findings, point out possible future directions and offer our conclusions; some of the details of our calculations can be found in the \textit{Appendix}.

	\section{Brief Review of the Effective Framework}
	\label{sec:review}
	
The original formulation of the effective field theory for inflation (EFTI) \cite{Cheung:2007st} provides a  general Lagrangian starting from the realisation that the inflaton background breaks time-reparametrisation invariance. Full diffeomorphism (diff) invariance can be restored including the Goldstone mode $\pi$ that non-linearly realises time diffs\footnote{This procedure is also known as the Stueckelberg trick (see e.g. \cite{ArkaniHamed:2002sp})}. The result is a theory that describes scalar perturbations around FRW and covers a vast range of single-field inflationary models (see \cite{Senatore:2010wk} for an extension of this approach).\\
There is a vast literature exploring the impact on cosmological observables resulting from adding extra ingredients to the single-field picture. The simplest next step is  to add scalar degrees of freedom \cite{Wands:2007bd}. Perhaps even more interesting a phenomenology results from considering the presence of extra spinning particles. This often comes at a price: severe bounds on the mass range, affecting most dramatically particles of spin-2 and higher according to  
\bea
\label{series}
s(s-1) \leq\frac{m^2}{H^2}< \left(s-\frac{1}{2}\right)^2 \quad;  \quad\frac{m^2}{H^2} \geq  \left(s-\frac{1}{2}\right)^2 ,
\eea
where the first expression corresponds to the principal series and the second one to the complementary series. We refer the reader to \cite{Arkani-Hamed:2015bza} and \cite{Lee:2016vti} where these conditions are explored at length in the inflationary context. It is also important to mention that interesting literature specifically dedicated to signatures of higher spin fields exists in \cite{MoradinezhadDizgah:2018pfo}. The set\footnote{Strictly speaking there is one more possibility  corresponding to partially massless theories:\\ $\frac{m^2}{H^2}=s(s-1)-t(t-1)$ with $t\leq s$ and $s,t=0,1,2,\dots$.
\noi Such set-ups come with additional gauge invariance \cite{Deser:1983mm,Deser:2001us} and therefore fewer degrees of freedom, hence the name. To date, no fully non-linear consistent \cite{Fasiello:2013woa,deRham:2013wv} theory of e.g. partially massless gravity has been identified. Remarkably, the latter formulation, if found, would be a premier candidate to solve the cosmological constant problem \cite{deRham:2014zqa}.} of inequalities in Eq.~(\ref{series}) stems from the familiar identification of particles as \textit{unitary} irreducible representation of the spacetime (de Sitter, in this case) isometry group \cite{Wigner:1939cj}. The most immediate consequence of Eq.~(\ref{series}) on observables is that extra spinning particles are heavy and have little to no impact on late time correlation functions: they typically decay during inflation, can only appear as internal Feynman lines and theirs is a small integrated-over-time effect.\\ 
There is, however, a way to go around these limitations and boost the effect of extra spinning fields. The key step is to exploit the fact that inflationary dynamics breaks de Sitter isometries: the inflaton background spontaneously breaks the Lorentz group down to the rotation group. Coupling extra fields directly with the inflaton will then relieve them from the strict conditions in Eq.~(\ref{series}). Intuitively speaking, ``non-minimal"\footnote{We shall use the phrasing ``non-minimal" throughout the text.  Following  \cite{Bordin:2018pca}, one might say, more accurately, that we are exploring the most general coupling of extra spinning particles to the (constant) inflaton foliation \textit{in the presence of an approximate shift symmetry for the inflaton}.} coupling to the inflaton field provides sufficient mixing with the non-decaying inflaton mode to keep extra spinning particles afloat, effectively allowing them to be light.\\
The procedure to obtain the most general Lagrangian mimics the one in \cite{Cheung:2007st}, with some important extra subtleties. The most immediate path stems from recognizing that fields should be identified as representations of the unbroken group (rotations): the Lagrangian one is after will then \textit{non-linearly} realise  the broken group. One starts with a field transforming in any tensor representation of the 3D rotation group and then embeds it in spacetime. For the case of a spin-2 particle, the resulting symmetric traceless tensor, propagating five degrees of freedom is embedded and coupled via \cite{Bordin:2018pca}:
\bea
\Sigma^{00}=\frac{\partial_i\pi\partial_j\pi}{(1+\dot{\pi})^2}\Sigma^{ij}\quad ;\quad  \Sigma^{0j}=-\frac{\p_i\pi}{(1+\dot{\pi})}\Sigma^{ij}\; ,
\eea
where $\pi(t,{\bf x})$ is the usual inflaton fluctuation, related to the standard curvature perturbation $\zeta$ via $\zeta=-H \pi$, at leading order. Equipped with the building blocks outlined above, and combining those with the usual ones from the EFTI, the authors of \cite{Bordin:2018pca} arrive at what will be our starting point: the quadratic and cubic inflationary Lagrangian including interactions between standard metric tensors fluctuations $\gamma_{ij}$ with those from a generic extra spin-2 field $\sigma_{ij}$, as well as self-interactions of $\sigma_{ij}$ itself \footnote{We do not consider here the contributions to the tensor bispectrum due to standard $\gamma$ self-interactions (see e.g. \cite{Maldacena:2002vr,Maldacena:2011nz}). In \textit{Section}~\ref{amplitude} we shall compare the latter  to those originating from the extra spin-2 field.}:
\begin{equation}
\mathcal{L}^{}_{\gamma+\sigma}=\mathcal{L}^{(2)}_{\text{free}}+\mathcal{L}^{(2)}_{\text{int}}+\mathcal{L}^{(3)}_{\text{int}}\,,
\end{equation}
where 
\begin{eqnarray}
&&\mathcal{L}^{(2)}_{\text{free}}=\frac{a^{3}}{8}\left[\left(\dot{\tilde\gamma}_{ij}\right)^2-\frac{1}{a^2}\left(\partial_{i}{\tilde\gamma}_{jk}\right)^{2}\right]+\frac{a^{3}}{4}\left[\left(\dot{\sigma}_{ij}\right)^2-\frac{c_{\sigma}^{2}}{a^2}\left(\partial_{i}\sigma_{jk}\right)^{2}-m_{\sigma}^2\left(\sigma_{ij}\right)^2\right]\,,\\\label{qm}
&&\mathcal{L}^{(2)}_{\text{int}}=\frac{a^{3}}{2}\,\rho\,\sigma^{ij}\dot{\tilde\gamma}_{ij} \,,\quad\quad\quad\mathcal{L}^{(3)}_{\text{int}}=-a^{3}\,\mu\left(\sigma_{ij}\right)^3\,,
\end{eqnarray}

\begin{figure}
\begin{center}
  \includegraphics[width=6cm]{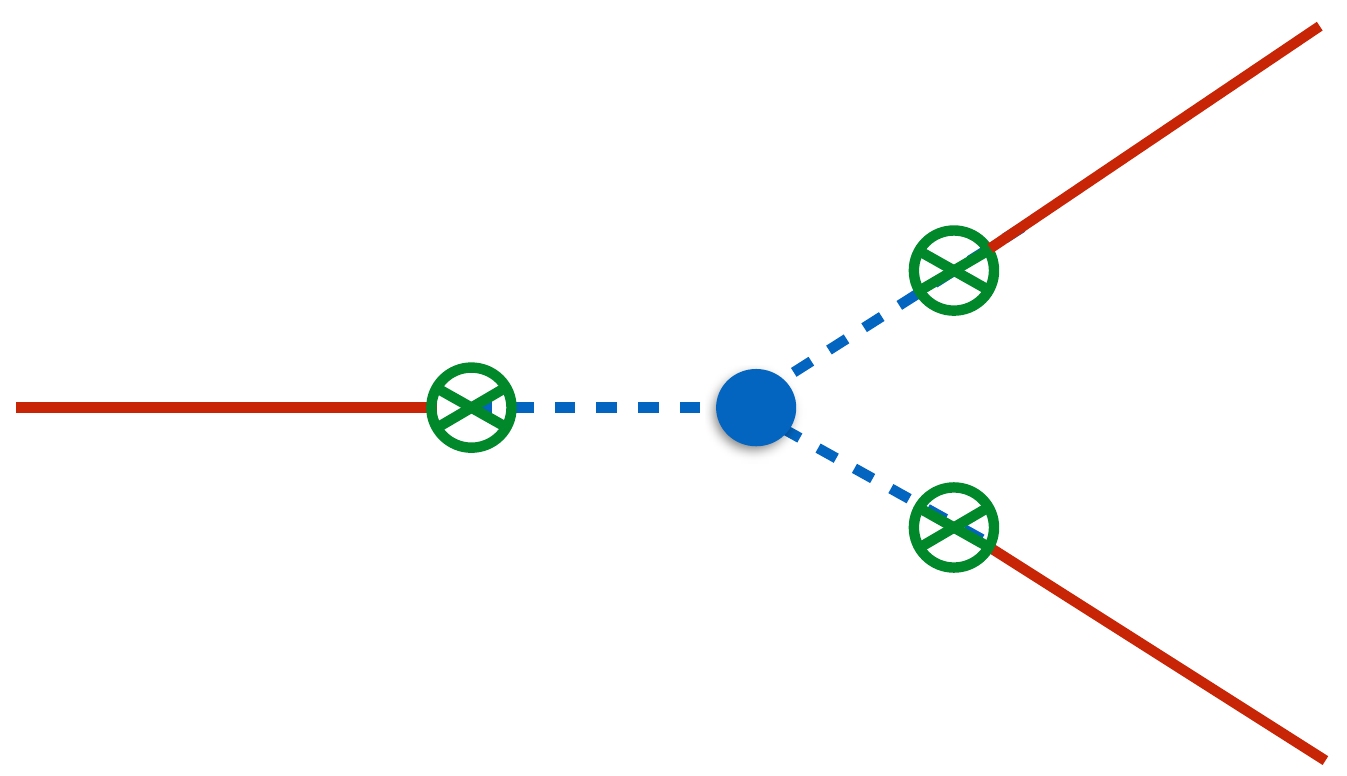}
  \caption
 {Diagrammatic representation of the contribution of $\sigma$ to the bispectrum of gravitational waves. Red/solid lines indicate propagators of $\gamma$, blue/dotted lines of $\sigma$. The blue vertex indicates the interaction given in $\mathcal{L}^{(3)}_{\text{int}}$, the green vertices correspond to the one from $\mathcal{L}^{(2)}_{\text{int}}$ (Eq.~(\ref{qm})).}
 \label{fig1}
\end{center}
\end{figure}

\noi with $c_{\sigma}$  the sound speed for the helicity-2 component of the extra spin-2 field. Note that we have canonically normalized metric fluctuations according to $\tilde{\gamma}_{ij}\equiv M_{\rm Pl} \gamma_{ij}$, where $\gamma_{ij}$ is the standard dimensionless metric perturbation. We will return to dimensionless quantities in  Eq.~(\ref{free}).
We take a sound speed equal to one for the tensor perturbations of the metric and $m_{\sigma}$ stands for the mass of the extra spin-2 field. The mass range of interest for us will be that corresponding to $0 \leq\nu\lesssim 1$, with $\nu=\sqrt{9/4-m_{\sigma}^2/H^2}$, which corresponds to interesting phenomenology for tensor non-Gaussianities and interpolates, as we shall see, between different shape profiles.\\
\noi The quadratic mixing and cubic self-interaction in (\ref{qm}) are the building blocks necessary to compute the diagram in Fig.~\ref{fig1}.\\

\noindent It is worth pointing out two notions at this stage:\\
-- the approach we are adopting here is a fully-fledged effective theory one, this is not akin to employing a specific parametrisation of fluctuations around FRW because here all operators compatible with the symmetries of the theory are automatically included\footnote{Corrections that are higher order in derivatives may be accounted for within the same formalism.};\\
-- the couplings to the inflaton fluctuations obtained via the construction in \cite{Bordin:2018pca} 
 are not arbitrary, they emerge from the fact that de Sitter isometries are non-linearly realised.

\section{Tensor bispectrum}
\label{sec:3}

 \noindent In this section we compute the tensor bispectrum mediated by the additional spin-2 field. Let us first report the solutions for the mode functions and for the tensor power spectrum. The tensor components of the spin-2 fields are expanded as
\begin{equation}\label{fexp1}
T_{ij}(\textbf{x},\tau)=\int\frac{d^{3}k}{(2\pi)^{3}}\,e^{i\textbf{k}\cdot\textbf{x}}\,\sum_{\lambda}\epsilon^{\lambda}_{ij}(\hat{k})T^{\lambda}_{\textbf{k}}(\tau)\,,
\end{equation}
 where $T$ is a place holder for $\gamma$ and $\sigma$. In what turns out to be a very convenient simplification, note that, upon requiring $\rho/H\ll 1$,  the mixing term $\mathcal{L}^{(2)}_{\text{int}}$ can be treated as a perturbation on top of $\mathcal{L}^{(2)}_{\text{free}}$. The very same $\rho/H$ ratio appears in late-time observables, starting with the tensor power spectrum. Similarly, we require $\mu/H\ll 1$ for $\mathcal{L}^{(3)}_{\text{int}}$ to be treated as a perturbation\footnote{The corresponding conditions in QsF \cite{Chen:2009zp} are $\dot{\theta}/H\ll1$ and $V^{'''}/H\ll 1$ , where $\dot{\theta}$ is the (constant)  turning angular velocity of the extra field, and $V=V(\sigma)$ is the potential for the same field.}. The mode-functions for $\sigma$ and $\gamma$ are then the solutions to the free-field equations    
\begin{eqnarray}
\label{free}
&&\gamma_{k}(\tau)=\frac{2 i H}{M_{\text{Pl}}\sqrt{2k^3}}\left(1+ik\tau\right)e^{-ik\tau}\,,\\  \label{solsigma}
&&\sigma_{k}(\tau)=\sqrt{\frac{\pi}{2}}H\left(-\tau\right)^{3/2}\,\mathcal{H}^{(1)}_{\nu}[-c_{\sigma} k \tau]\,,
\end{eqnarray}
where $\nu\equiv\sqrt{9/4-(m_{\sigma}^2/H^2)}$ and $\mathcal{H}^{(1)}_{\nu}$ is the Hankel function of the first kind. The power spectrum of gravitational waves receives a contribution from the mixing with the extra spin-2 field
\begin{eqnarray}\label{tps}
&&P_{\gamma}(k)=\frac{4H^2}{M_{\text{Pl}}^2\,k^3}\left[1+\frac{\mathcal{C}_{\gamma}(\nu)}{c_{\sigma}^{2\nu}}\left(\frac{\rho}{H}\right)^2\right]\equiv\frac{4H^2}{M_{\text{Pl}}^2\,k^3}\left[1+\alpha_{\gamma}\right]\,,
\end{eqnarray}
where the function $C_{\gamma}(\nu)$ can be found in \cite{Bordin:2018pca} and applies to the case of finite mass for $\sigma$. The quantity $C_{\gamma}(\nu)$ encapsulates the effect of the quadratic $\gamma-\sigma$ interaction on the tensor power spectrum. We stress here that, for the $\nu$ range of interest, this quantity as well as its scalar sector counterpart $\mathcal{C}_{\zeta}(\nu)$ in Eq.~(\ref{pssca}) are of order one to ten.

\noi Upon using the parametrization for the power spectrum in the RHS of Eq.~(\ref{tps}), one finds for the tilt
\bea 
n_{T}=-2\epsilon+\frac{\dot{\alpha}_{\gamma}}{H(1+\alpha_{\gamma})}\,.
\eea
One  therefore needs to allow for a time dependence in $c_{\sigma}$ or $\rho$ in order to support a blue tensor spectrum \footnote{As discussed in the introduction, there exist a number of mechanisms for generating a blue-tilted PGW spectrum. Another interesting class of such models relies on the breaking of space diffeomorphism invariance during inflation, resulting in an effective mass for tensor fluctuations (see e.g. \cite{cT}).}. We do not expect a time-dependent sound speed to qualitatively change, for example, the shape-function profile of non-Gaussianities and it is also with this in mind that we will provide in \textit{Section} \ref{lrf} our results for the LISA response function for a bispectrum momentum dependence that, as is the case for the model under consideration, interpolates between the local and the equilateral shapes (see \textit{Section}~\ref{shapefunction}). Naturally, one may look for a large primordial tensor signal already at CMB scales. 
Planned future probes, such as the LiteBIRD satellite, will measure polarization anisotropies over the full sky. LiteBIRD will constrain the tensor to scalar ratio down to $\Delta r=0.001$ and improve the constraints on $f^{\rm tensor}_{\rm nl}$ \cite{Matsumura:2013aja}. CMB-S4 experiments \cite{Abazajian:2016yjj} will also deliver substantial improvements on current bounds.\\

\noindent We now move on to presenting our results for the contribution to the gravitational waves bispectrum represented by the diagram in Fig.~(\ref{fig1}), and refer the reader to Appendix~\ref{AppendixA} for further related details. The $\sigma$-mediated contribution reads %
\begin{eqnarray}\label{eee}
\langle \gamma^{\lambda_{1}}_{\textbf{k}_{1}} \gamma^{\lambda_{2}}_{\textbf{k}_{2}} \gamma^{\lambda_{3}}_{\textbf{k}_{3}}\rangle_{(\sigma)}&=& (2\pi)^3\delta^{(3)}(\textbf{k}_{1}+\textbf{k}_{2}+\textbf{k}_{3})\mathcal{A}^{\lambda_{1}\lambda_{2}\lambda_{3}}\,B_{(\sigma)}(k_{1},k_{2},k_{3})\,,
\end{eqnarray}
where 
\bea
\label{helicities}
\mathcal{A}^{\lambda_{1}\lambda_{2}\lambda_{3}}\equiv\epsilon^{\lambda_{1}}_{\alpha\beta}(-\hat{k}_{1})\epsilon^{\lambda_{2}}_{\beta\gamma}(-\hat{k}_{2})\epsilon^{\lambda_{3}}_{\gamma\alpha}(-\hat{k}_{3})\; ,
\eea
and
\begin{eqnarray}\label{fullb}
B_{(\sigma)}(k_{1},k_{2},k_{3})&=&\frac{12\pi^3}{k_{1}^{4}k_{2}k_{3}}\frac{\mu}{H}\frac{\rho^{3}}{M_{\text{Pl}}^{3}}\mathcal{F}_{\nu c_{\sigma}}(k_{1},k_{2},k_{3})+\text{5\,perms}\,.
\end{eqnarray}
Here $\mathcal{F}_{\nu c_{\sigma}}$ is a lengthy function of the momenta and it depends on the parameters $\nu$ and $c_{\sigma}$, see  Eqs.~(\ref{A123general})-(\ref{C123general}). \noi
It is important to point out that the tensor bispectrum, unlike its scalar counterpart, also depends on the orientation of the 3-momentum vectors $\vec k_i$, and not merely on their norm. As a result, when performing calculations it will be necessary  to choose a reference orientation for the $k_i$ triangle sides. For simplicity, we avoid including the explicit dependence on the directions in $B_{(\sigma)}$ and express it only in terms of the size $k_i$ of the three momenta. Also, we point out that the bispectrum dependence on chirality has been extracted in the overall factor $\mathcal{A}^{\lambda_1\lambda_2\lambda_3}$, and there is no intrinsic chirality dependence in the bispectrum $B_{(\sigma)}\,$.

\subsection{Amplitude}
\label{amplitude}

As can be anticipated by means of the analogy with the QsF case, the contributions to the tensor bispectrum we are after will lead to a non-zero amplitude both in the equilateral ($k_1\sim k_2\sim k_3$) and in the squeezed limit ($k_3\ll k_2\sim k_3$) for most values of the mass $m_{\sigma}$. We find it useful then to provide below the complete expressions for the bispectrum (\ref{fullb}) in the squeezed  and equilateral  configurations 
\begin{eqnarray}\label{eqqqqq}
&& B_{(\sigma)}(k_{1},k_{2},k_{3})|_{k_{1}\sim k_{2}\gg k_{3}}\simeq\frac{3\pi^2\cdot 2^{3+\nu}}{k_{1}^{7/2-\nu}k_{2}^{}\,k_{3}^{3/2+\nu}}\, \frac{\mu}{H}\frac{\rho^3}{M_{\text{Pl}}^{3}}\, s^{\text{sq}}(\nu, c_{\sigma})\,,\\\label{eqqqq}
&& B_{(\sigma)}(k_{1},k_{2},k_{3})|_{k_{1}\sim k_{2}\sim k_{3}}=\frac{9 \pi^3\cdot 2^3}{k_{1}^{6}} \frac{\mu}{H}\frac{\rho^3}{M_{\text{Pl}}^{3}}\,s^{\text{eq}}(\nu, c_{\sigma})\,,
\end{eqnarray}
where 
\begin{eqnarray}\label{sq1}
s^{\text{sq}}(\nu, c_{\sigma})&\equiv&\frac{\Gamma[\nu]}{c_{\sigma}^{\nu}}\int_{-\infty}^{0}dx_{1}\int_{-\infty}^{x_{1}}dx_{2}\int_{-\infty}^{x_{2}}dx_{3}\Big\{(-x_{1})^{-1/2}(-x_{2})^{1/2-\nu}(-x_{3})^{-1/2}\sin(-x_{1})\nonumber\\&&\times\text{Im}\left[\mathcal{H}_{\nu}^{(1)}(-c_{\sigma}x_{1})\mathcal{H}_{\nu}^{(2)}(-c_{\sigma}x_{2})\right]\text{Im}\left[\mathcal{H}_{\nu}^{(1)}(-c_{\sigma}x_{2})\mathcal{H}_{\nu}^{(2)}(-c_{\sigma}x_{3})\,e^{ix_{3}}\right]\nonumber\\&&+(-x_{1})^{-1/2}(-x_{2})^{-1/2}(-x_{3})^{1/2-\nu}\sin(-x_{1})\sin(-x_{2})\nonumber\\&&\times\text{Im}\left[\mathcal{H}_{\nu}^{(2)}(-c_{\sigma}x_{1})\mathcal{H}_{\nu}^{(2)}(-c_{\sigma}x_{2})\left(\mathcal{H}_{\nu}^{(1)}(-c_{\sigma}x_{3})\right)^2\right]\nonumber\\&&\times\int_{-\infty}^{0}dy_{4}(-y_{4})^{-1/2}\text{Re}\left[\mathcal{H}_{\nu}^{(1)}(-c_{\sigma}y_{4})\,e^{-iy_{4}}\right]
\Big\}\,,
\end{eqnarray}
and
\begin{eqnarray}\label{seq1}
s^{\text{eq}}(\nu, c_{\sigma})&\equiv&\int_{-\infty}^{0}dx_{1}\int_{-\infty}^{x_{1}}dx_{2}\int_{-\infty}^{x_{2}}dx_{3}\int_{-\infty}^{x_{3}}dx_{4}\Big\{\sqrt{\frac{x_{2}}{x_{1}x_{3}x_{4}}}\text{Im}\left[\mathcal{H}_{\nu}^{(1)}(-c_{\sigma}x_{1})\mathcal{H}_{\nu}^{(2)}(-c_{\sigma}x_{2})\right]\nonumber\\&&\times\text{Im}\left[\mathcal{H}_{\nu}^{(2)}(-c_{\sigma}x_{2})\mathcal{H}_{\nu}^{(1)}(-c_{\sigma}x_{4})\,e^{-ix_{4}}\right]\text{Im}\left[\mathcal{H}_{\nu}^{(1)}(-c_{\sigma}x_{2})\mathcal{H}_{\nu}^{(2)}(-c_{\sigma}x_{3})\,e^{ix_{3}}\right]\sin(-x_{1})\nonumber\\&&+\sqrt{\frac{x_{3}}{x_{1}x_{2}x_{4}}}\sin(-x_{1})\sin(-x_{2})\text{Im}\left[\mathcal{H}_{\nu}^{(2)}(-c_{\sigma}x_{1})\mathcal{H}_{\nu}^{(2)}(-c_{\sigma}x_{2})\left(\mathcal{H}_{\nu}^{(1)}(-c_{\sigma}x_{3})\right)^2\right]\nonumber\\&&\times\text{Im}\left[\mathcal{H}_{\nu}^{(2)}(-c_{\sigma}x_{3})\mathcal{H}_{\nu}^{(1)}(-c_{\sigma}x_{4})\,e^{-ix_{4}}\right]+\sqrt{\frac{x_{4}}{x_{1}x_{2}x_{3}}}\sin(-x_{1})\sin(-x_{2})\sin(-x_{3})\nonumber\\&&\times\text{Im}\left[\left(\mathcal{H}_{\nu}^{(1)}(-c_{\sigma}x_{4})\right)^{3}\mathcal{H}_{\nu}^{(2)}(-c_{\sigma}x_{1})\mathcal{H}_{\nu}^{(2)}(-c_{\sigma}x_{2})\mathcal{H}_{\nu}^{(2)}(-c_{\sigma}x_{3})\right]
\Big\}\,.
\end{eqnarray}

\noindent Let us now compute the bispectrum amplitude, $f_{\text{nl}}$, which we define as\footnote{We choose this definition for the amplitude in order to conform with Planck \cite{planckng}. An alternative definition for  estimating the size of tensor non-Gaussianity is 
 $B_{\sigma}/P_{\gamma}^{2}$, which only makes use of the properties of tensor modes with no reference to scalar fluctuations.
 }
\begin{equation}\label{newdef}
B_{(\sigma)}(k_{1},k_{2},k_{3})=f_{\text{nl}}\,\left(\frac{H^{2}}{4\epsilon\,M_{\text{Pl}}^{2}}\right)^{2}\,F(k_{1},k_{2},k_{3})\,,
\end{equation}
where (from Eqs.~(\ref{eqqqqq}), (\ref{eqqqq})):
\begin{equation}\label{Alll}
F(k_{1},k_{2},k_{3})=\begin{cases}
    \frac{1}{k_{1}^{6}}, & \,\, \text{for $k_{1}\sim k_{2}\sim k_{3}$}.\\
  \frac{1}{k_{1}^{9/2-\nu}\,k_{3}^{3/2+\nu}}, & \,\, \text{for $k_{1}\sim k_{2}\gg k_{3}$}.
  \end{cases}
\end{equation}
\noindent We will compute the bispectrum amplitudes for $\nu=1$. We choose this particular value because we find it particularly instructive in that the corresponding shape-function exhibits a non-zero component both in the local and equilateral configurations. It is well-known that, upon employing massive fields, one may indeed interpolate between these two standard templates. We shall see this more explicitly in \textit{Section} \ref{shapefunction}.\\

\begin{figure}[h]
    \hspace{-10mm}
  \includegraphics[width=86mm]{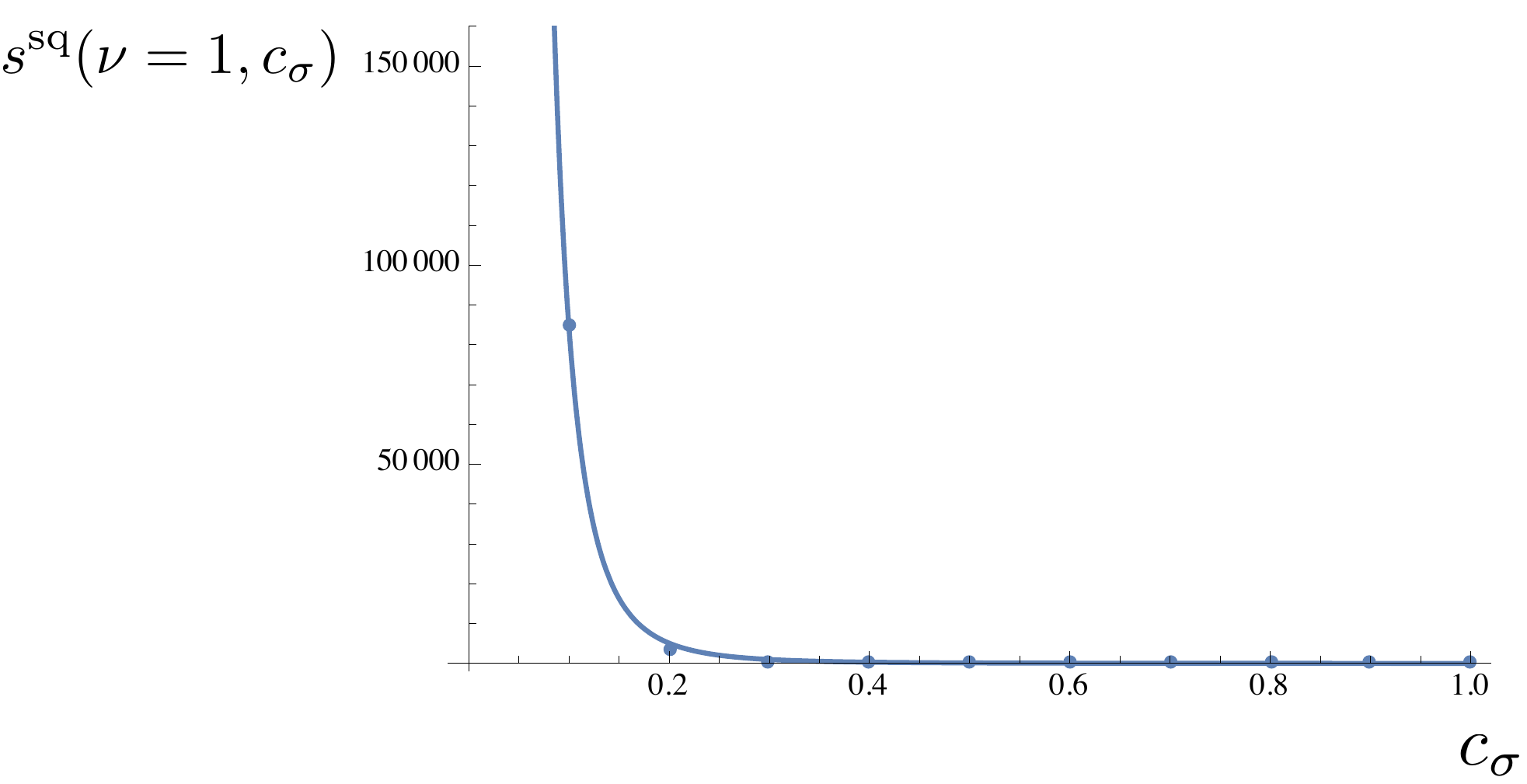}
  \includegraphics[width=87mm]{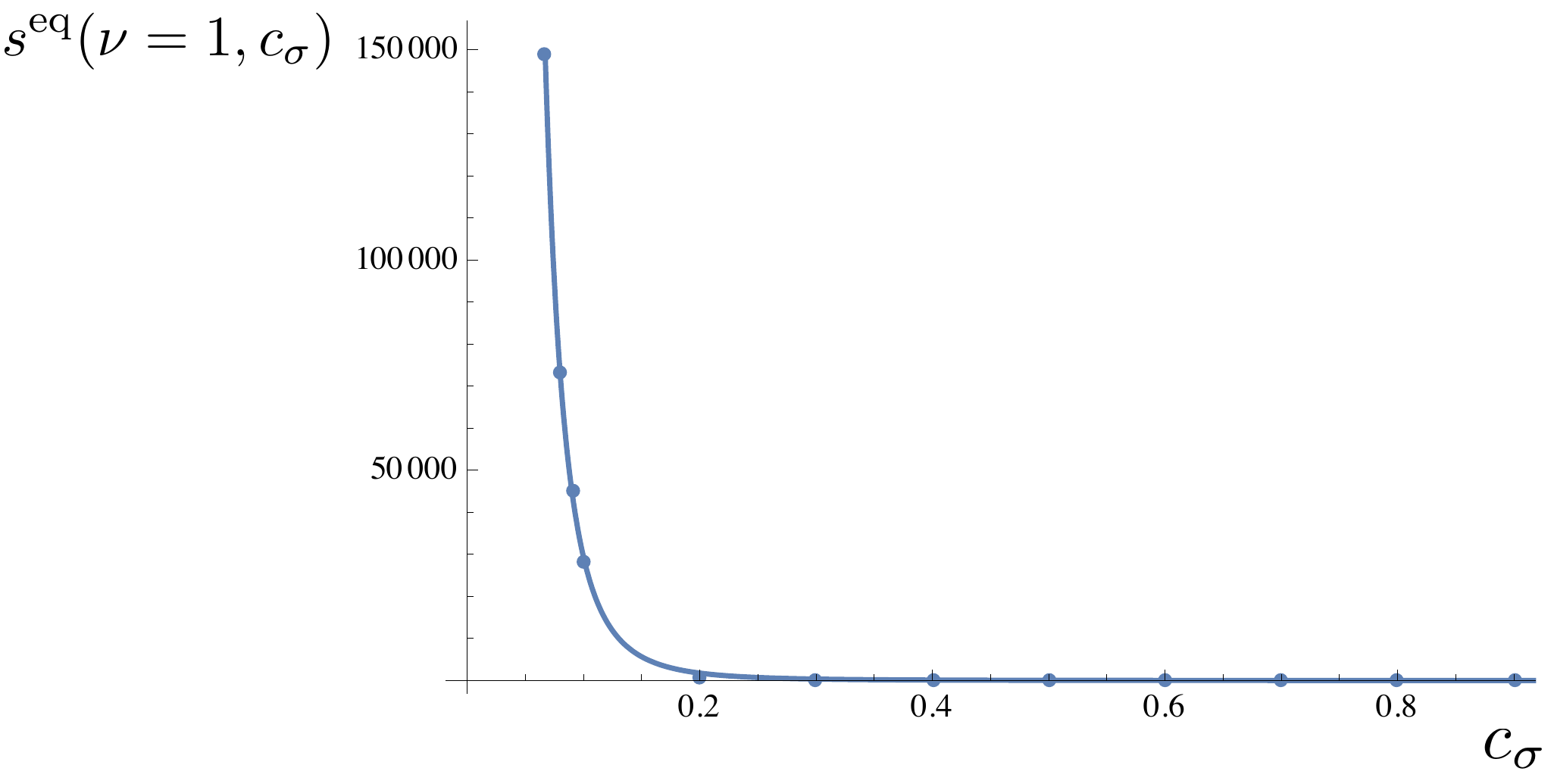}
  \caption
 {{\bf(Left panel)} Plot of numerical values of the coefficient $s^{\text{sq}}(\nu,c_{\sigma})$ defined in Eq. (\ref{sq1}), as a function of $c_{\sigma}$ and for $\nu=1$. The solid line corresponds to the fit in Eqs.~(\ref{fits}).
{\bf(Right panel)} Numerical values for $s^{\text{eq}}(\nu,c_{\sigma})$ (Eq.~(\ref{seq1})) as a function of the sound speed for $\nu=1$, with the solid line representing the functional dependence in Eq.~(\ref{fits}).}
 \label{fig2}
\end{figure}

\begin{figure}[tbp]
    \hspace{-11mm}
  \includegraphics[width=72mm]{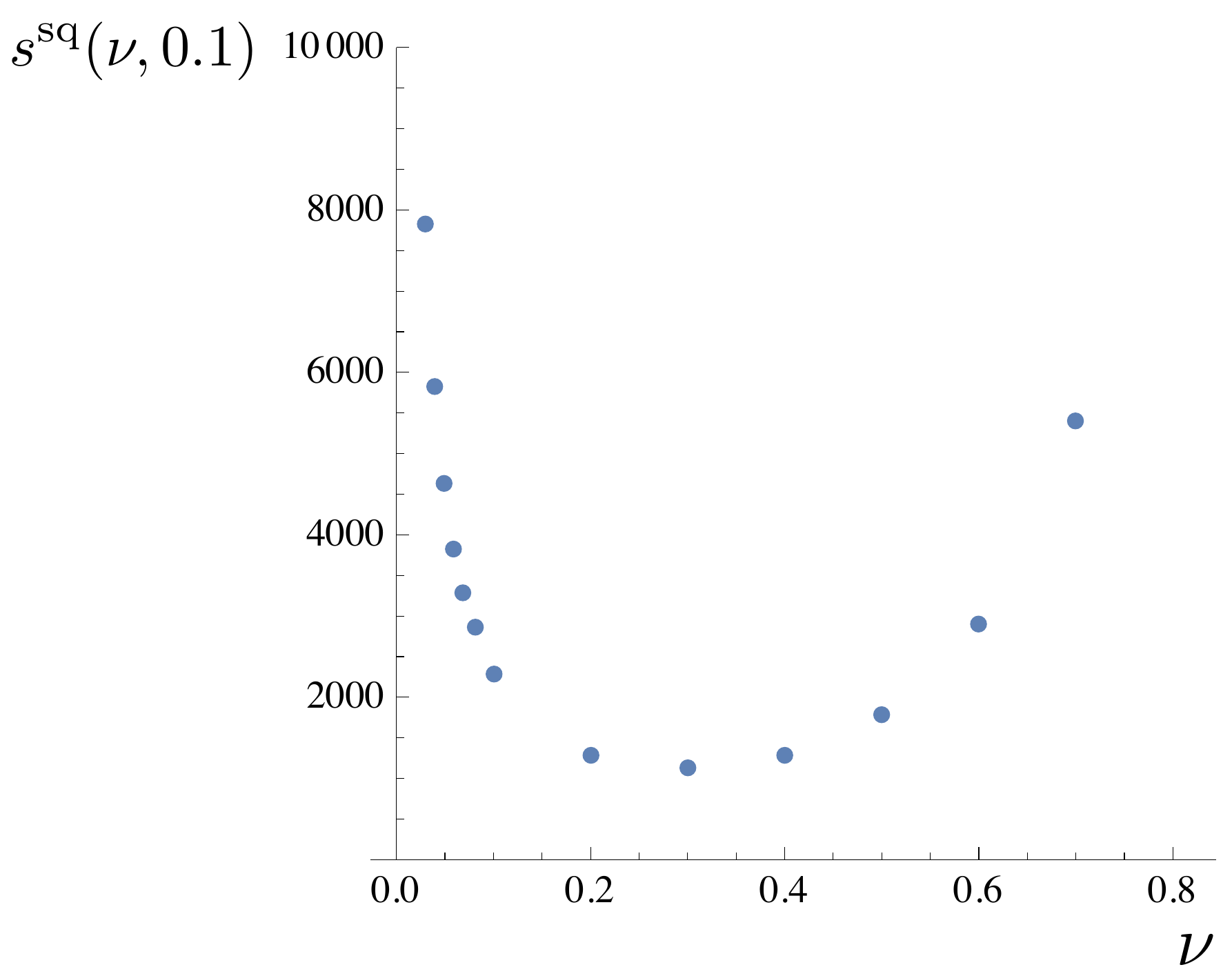}
  \hspace{5mm}
  \includegraphics[width=92mm]{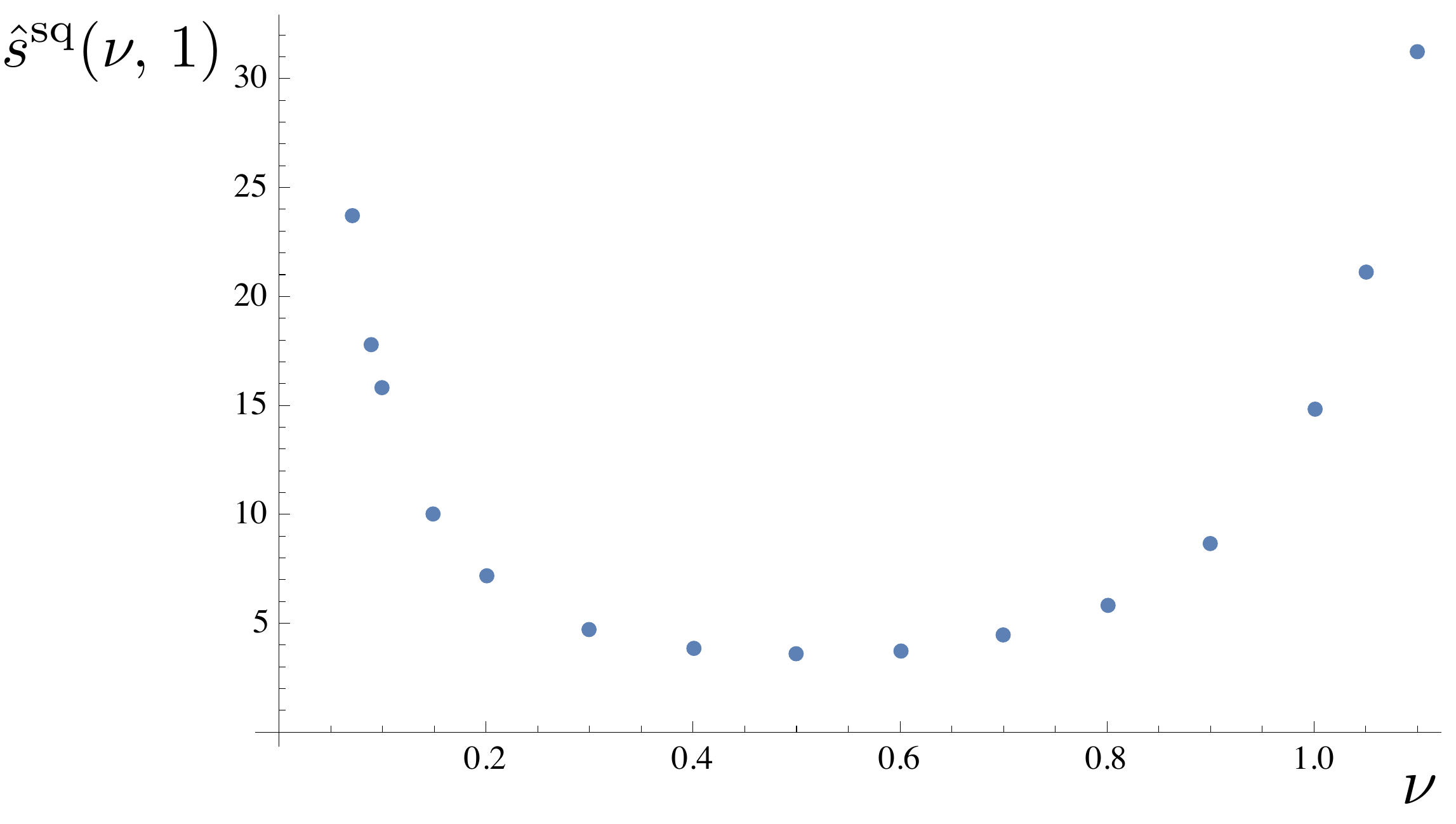}
  \caption
 {{\bf(Left panel)} Numerical values of $s^{\text{sq}}(\nu,c_{\sigma})$ for fixed $c_{\sigma}=0.1$. 
{\bf(Right panel)} Numerical values of $\hat{s}^{\text{sq}}(\nu,c_{\sigma})\equiv (3\pi^2/2^{3-\nu})\, s^{\text{sq}}(\nu,c_{\sigma})$ for $c_{\sigma}=1$. Here we rescaled the coefficient $s^{\text{sq}}(\nu,c_{\sigma})$ for the sake of comparison with \cite{Chen:2009zp} (see Fig.~9 therein).}
 \label{fig3}
\end{figure}

\begin{figure}[tbp]
    \hspace{-14mm}
  \includegraphics[width=88mm]{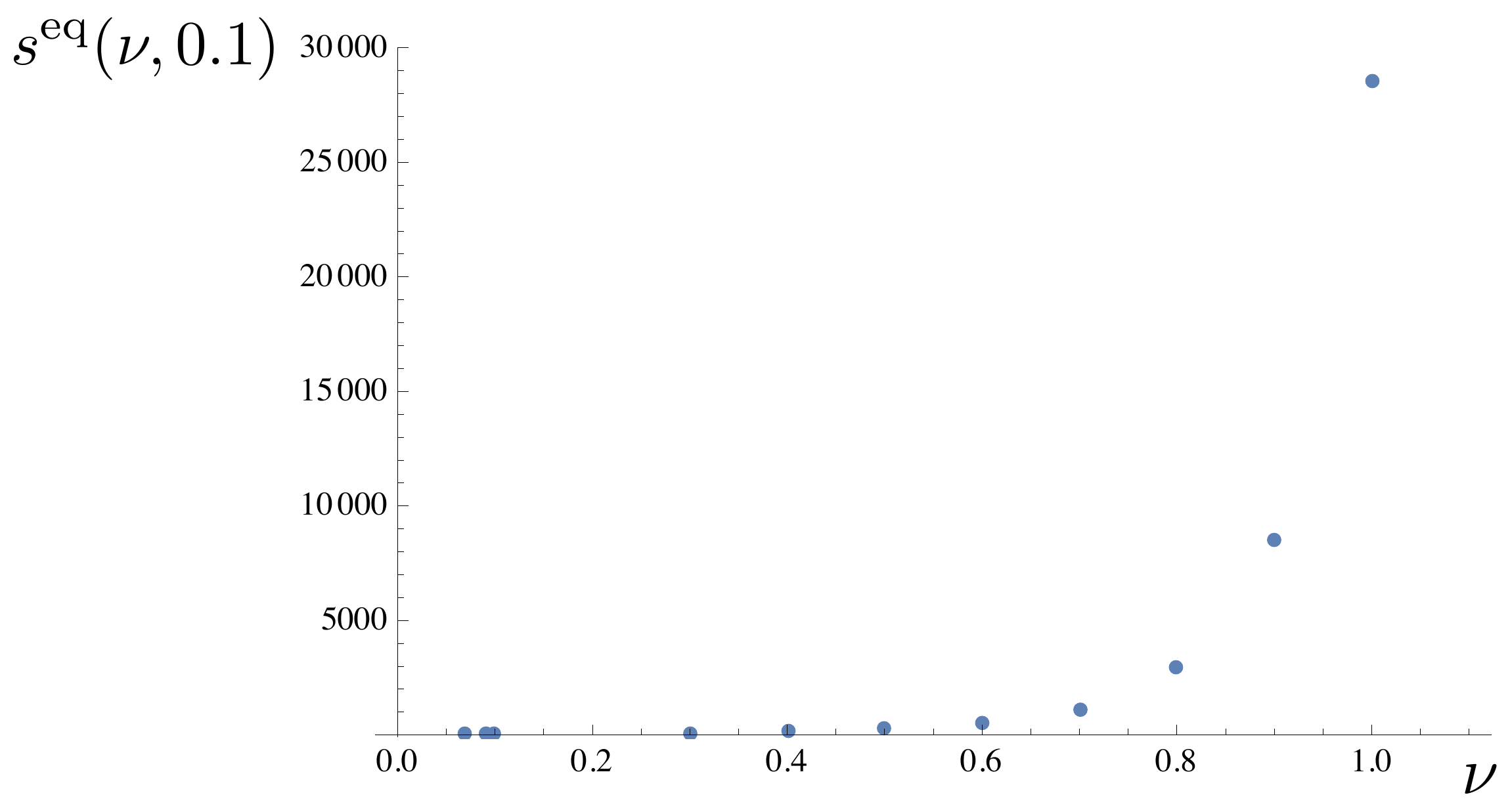}
  \includegraphics[width=92mm]{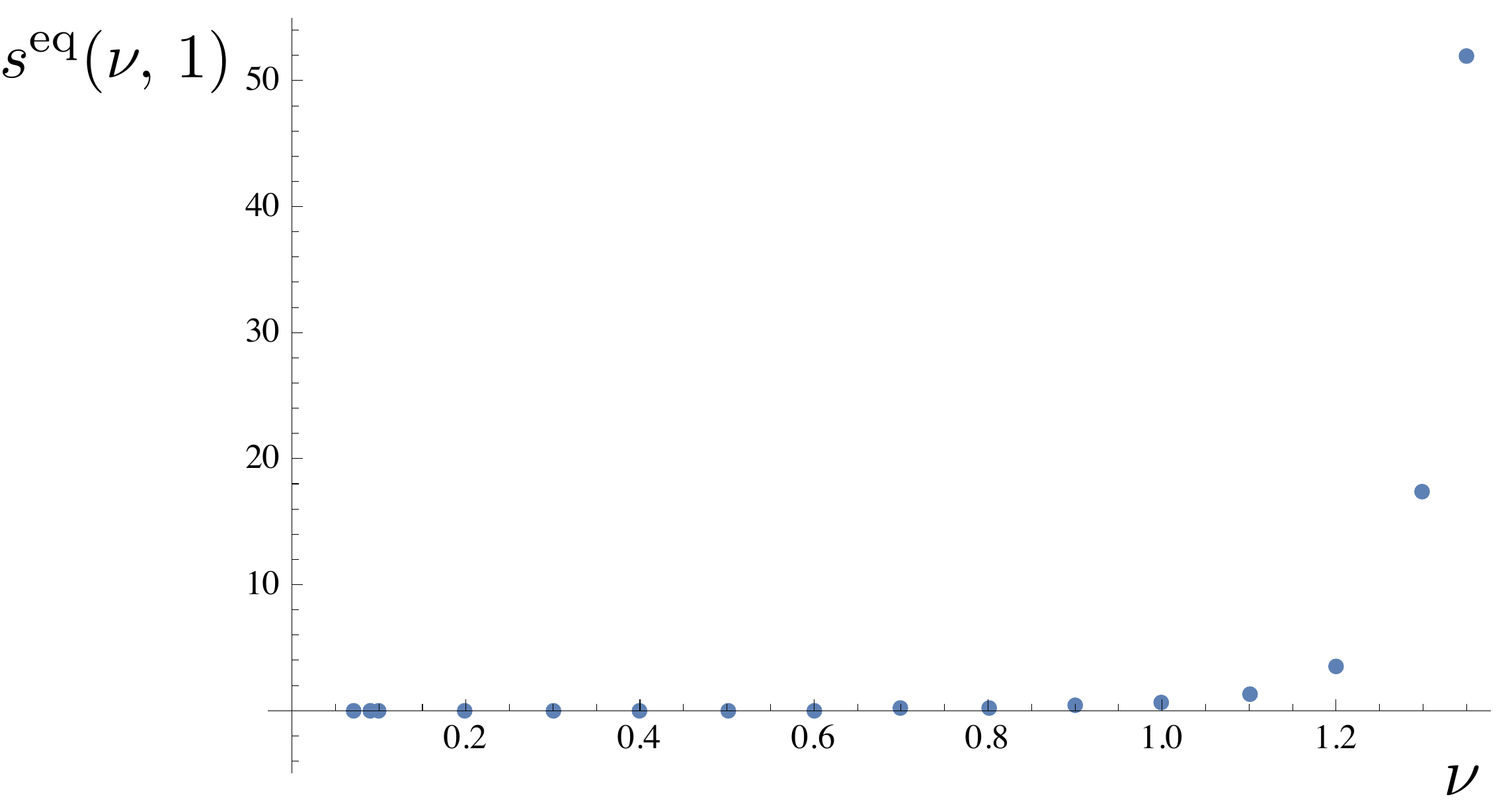}
  \caption
 {{\bf(Left panel)} Numerical values of $s^{\text{eq}}(\nu,c_{\sigma})$ for fixed $c_{\sigma}=0.1$. 
{\bf(Right panel)} Numerical values of $s^{\text{eq}}(\nu,c_{\sigma})$ for fixed $c_{\sigma}=1$.}
\label{fig4}
\end{figure}

\noi We report in Fig.~(\ref{fig2}) our numerical results for the functional behaviour of $s^{\text{sq}}(c_{\sigma},\nu)$ and $s^{\text{eq}}(c_{\sigma},\nu)$ for $\nu=1$, which are consistent with the power laws
\begin{eqnarray}\label{fits}
s^{\text{sq}}(1, c_{\sigma})\simeq \frac{7.4}{c_{\sigma}^{4.06}}\,,\quad\quad s^{\text{eq}}(1, c_{\sigma})\simeq \frac{2.5}{c_{\sigma}^{4.05}}\,.
\end{eqnarray}
Figs.~(\ref{fig3}) and (\ref{fig4}) show the behaviour of, respectively, $s^{\text{sq}}$ and $s^{\text{eq}}$, as a function of $\nu$ for fixed values of the sound speed. For $c_{\sigma}=1$ our numerical results are in agreement with the calculations in the scalar sector of QsF inflation \cite{Chen:2009zp}.\\

\noindent The scalar power spectrum is given by \cite{Bordin:2018pca}
\begin{equation}\label{pssca}
P_{\zeta}(k)=\frac{H^2}{4\epsilon M_{\text{Pl}}^2\,k^3}\left[1+\frac{\mathcal{C}_{\zeta}(\nu)}{\epsilon\,c_{0}^{2\nu}}\left(\frac{\rho}{H}\right)^2\right]\equiv\frac{H^2}{4\epsilon M_{\text{Pl}}^2\,k^3}\left[1+\alpha_{\zeta}\right]\,,
\end{equation}
where $c_{0}$ is the sound speed for the helicity zero component of $\sigma$. The condition $\rho\ll c_{0}\sqrt{\epsilon}H$ ensures that radiative corrections to $m_{\sigma}$ remain small \cite{Bordin:2018pca}. For values of $\nu$ in the vicinity of 1, this implies $\alpha_{\zeta}\ll 1$ and therefore a scalar power spectrum dominated by vacuum fluctuations, $P_{\zeta}\simeq H^{2}/(4\epsilon M_{\text{Pl}}^{2}k^{3})$. On the other hand, the tensor power spectrum, (\ref{tps}), may be dominated either by the sourced or by the vacuum contributions (or the two may be comparable to one another) depending on parameters choice. One can therefore employ the following relation in Eq.~(\ref{newdef}): $H^{2}/(4\epsilon\,M_{\text{Pl}}^{2})=(1/r)\cdot(1+\alpha_{\gamma})\,4\,H^{2}/M_{\text{Pl}}^{2}\;$, where the tensor to scalar ration $r$ is defined as usual by $r\equiv \mathcal{P}_{\gamma}/\mathcal{P}_{\zeta}$ .\\
\noindent Let us consider the equilateral limit first. Using Eqs.~(\ref{eqqqq}) and (\ref{fits}), the definition in (\ref{newdef}) gives
\begin{equation}\label{ppp1}
f_{\text{nl}}^{\text{eq}}\simeq r^2\cdot \left(\frac{2.5\cdot9\cdot\pi^3}{2}\right)\left(\frac{\mu}{H}\right)\left(\frac{\rho}{H}\right)^3\left(\frac{M_{\text{Pl}}}{H}\right)\frac{1}{c_{\sigma}^{4}\,(1+\alpha_{\gamma})^2}\,,
\end{equation}

\noindent Let us analyze the three cases: $\alpha_{\gamma}\ll 1$, $\alpha_{\gamma}\gg 1$ and $\alpha_{\gamma}\simeq 1$, using the relation $M_{\text{Pl}}\sqrt{\epsilon}/H\simeq10^{3}$ (from the amplitude of the scalar power spectrum), together with $r=16\,\epsilon\,(1+\alpha_{\gamma})$ (from the definition of tensor-to-scalar ratio combined with Eq.~(\ref{tps})). From (\ref{tps}) and for $\nu=1$ one also finds: $\alpha_{\gamma}(1+\alpha_{\gamma})=(10/16)(r/c_{\sigma}^{2})(\rho/(\sqrt{\epsilon}H))^2$. These relations are to be combined with the constraints $\rho/(\sqrt{\epsilon}H)\ll 1$ and $\mu/H\ll 1$ .\\
\noi Equation~(\ref{ppp1}) can be rewritten\footnote{Note that, to facilitate the comparison with the standard single-field slow-roll result, we isolate in the expressions below the factor $r^2$.} as  
\begin{equation}
f_{\text{nl}}^{\text{eq}}\simeq \left[4.4\times 10^{4}\left(\frac{\mu}{H}\right)\left(\frac{1}{c_{\sigma}}\right)\left(\frac{1}{r^{1/2}}\right)\left(\frac{\alpha_{\gamma}}{1+\alpha_{\gamma}}\right)^{3/2}\right]r^{2}\,.
\end{equation}
As an example, a tensor-to-scalar ratio near the current bound of $0.06$ \cite{Akrami:2018odb} and a value of $\mu/H= 0.5$ would yield:\\

\noindent $(i)$ $\alpha_{\gamma}=\mathcal{O}(10^{-1})$:
\begin{equation}\label{amp1}
f_{\text{nl}}^{\text{eq}}\simeq \begin{cases}
    \frac{6957}{\sqrt{r}}\,r^2\simeq 102 & \,\, \text{for $c_{\sigma}=0.1$}\,;\\
    \frac{12060}{\sqrt{r}}\,r^2\simeq 177, & \,\, \text{for $c_{\sigma}=0.05$}\,;\\
    \frac{60302}{\sqrt{r}}\,r^2\simeq 886 & \,\, \text{for $c_{\sigma}=0.01$}\,.
  \end{cases}
\end{equation}

\noindent $(ii)$ $\alpha_{\gamma}=\mathcal{O}(1)$:
\begin{equation}\label{amp2}
f_{\text{nl}}^{\text{eq}}\simeq \begin{cases}
    \frac{77782}{\sqrt{r}}\,r^2\simeq 1143 & \,\, \text{for $c_{\sigma}=0.1$}\,;\\
    \frac{155563}{\sqrt{r}}\,r^2\simeq 2286, & \,\, \text{for $c_{\sigma}=0.05$}\,
    \\
    \frac{777817}{\sqrt{r}}\,r^2\simeq 11431 & \,\, \text{for $c_{\sigma}=0.01$}\,.
  \end{cases}
\end{equation}

\noindent $(iii)$ $\alpha_{\gamma}=\mathcal{O}(10)$: in this case, only values $c_{\sigma}< 0.02$ are compatible with the above conditions on the parameter space
\begin{equation}
\label{amp3}
f_{\text{nl}}^{\text{eq}}\simeq 
    \frac{2\cdot 10^{6}}{\sqrt{r}}\,r^2\simeq  28026  \quad \text{for \; $c_{\sigma}=0.01$} \;.
\end{equation}
Using Eqs.~(\ref{eqqqqq}), (\ref{fits}), and (\ref{newdef}), one finds slightly smaller values for the amplitude in the squeezed limit 
\begin{equation}
f_{\text{nl}}^{\text{sq}} \simeq 0.6\,f_{\text{nl}}^{\text{eq}}\,.
\end{equation}

\noindent These values of the amplitude can be compared\footnote{We note here that, unlike the case of $f^{\rm eq}_{\rm nl}$ ,  our definition of $f^{\rm sq}_{\rm nl}$ differs from the one associated with the local template typical of single-field slow-roll inflation.

} with those due to bispectrum contributions from self-interactions of $\gamma$, which are also the only ones present in the minimal single-field inflationary models. One finds, for example \cite{Maldacena:2002vr,Maldacena:2011nz} (see e.g. also \cite{Agrawal:2018gzp}): 
\begin{eqnarray}\label{ampsfsr}
&&f_{\text{nl},\text{standard}}^{\lambda_{1}\lambda_{2}\lambda_{3}(\text{eq})}\simeq\begin{cases}
  0.9\, r^2, & \,\, \text{for $\,\,\lambda_{1}\lambda_{2}\lambda_{3}=RRR,\,LLL$}\,,\\
     0.01\, r^2, & \,\, \text{for $\,\,\lambda_{1}\lambda_{2}\lambda_{3}=RRL,\,LLR$}\,,
  \end{cases}
  \\
  \label{ampsfsr1}
&&f_{\text{nl},\text{standard}}^{\lambda_{1}\lambda_{2}\lambda_{3}(\text{sq})}\simeq 0.2\,r^2,\quad\quad\,\,\, \text{for $\,\,\lambda_{1}\lambda_{2}\lambda_{3}=RRR,\,LLL,\,RRL,\,LLR\,,$}
\end{eqnarray}
for $\alpha_{\gamma}\leq 1$. For $\alpha_{\gamma}\gg 1$ the amplitudes in Eqs.~(\ref{ampsfsr})-(\ref{ampsfsr1}) are suppressed by a factor $\sim \alpha_{\gamma}^{-2}$. Notice that in (\ref{ampsfsr})-(\ref{ampsfsr1}) we redefined the amplitude of the bispectrum to include the information on polarization. In other words, the amplitudes in (\ref{ampsfsr})-(\ref{ampsfsr1}) are to be compared with $f_{\text{nl}}^{\lambda_{1}\lambda_{2}\lambda_{3}(\text{eq/sq})}\equiv \mathcal{A}^{\lambda_{1}\lambda_{2}\lambda_{3}}f_{\text{nl}}^{(\text{eq/sq})}$ where, from its definition in Eq.~(\ref{helicities}) (see also (\ref{hel})) one finds the following values for $\mathcal{A}^{\lambda_{1}\lambda_{2}\lambda_{3}}$:
\begin{center}
\begin{table}[ht]
\hspace{0.15cm}
\begin{tabular}{|l|l|l|}
\hline
$\quad$ & \thead{\textbf{Squeezed limit ($k_{1}\sim k_{2}\gg k_{3}$)}} & \thead{\textbf{Equilateral limit ($k_{1}\sim k_{2}\sim k_{3}$)}}                     \\ \hline
\thead{\textbf{RRR, LLL}} & 1/4 &   27/64                    \\ \hline
\thead{\textbf{RRL, LLR}}  & 1/4 &     3/64                  \\ 
\hline
\makecell{\textbf{LRR, RLL,}\\ \textbf{RLR, LRL}} & $(1/16)\cdot(k_{3}^{2}/k_{2}^{2})$  &  3/64\\ 
\hline
\end{tabular}
\caption{Values of the function $\mathcal{A}^{\lambda_{1}\lambda_{2}\lambda_{3}}$ for various helicities combinations.}
\label{table1}
\end{table}
\end{center}
We have found that $c_{\sigma}$ values smaller than unity lead to an amplitude for the bispectrum sourced by the extra spin-2 field that can be several orders of magnitude larger than the one predicted in standard single-field slow-roll
inflation.

\subsection{Shape-Function}\label{shapefunction}
We now move on to study the dependence of the tensor bispectrum on the $(k_1,k_2,k_3)$ configuration. The observable is a function of the norm of the wavenumbers and one can set, without loss of generality, the norm of one such wavenumber to one. Momentum conservation dictates that the three external momenta of the bispectrum form a closed triangle and, depending on the configuration probed, one refers to e.g. the aforementioned equilateral and squeezed cases. In  analogy with the scalar QsF case, we expect our bispectrum to peak (whilst still exhibiting a non-zero value in the equilateral region) in the squeezed limit for $\nu \gtrsim 1$ and to display a plateau like profile for smaller value of $\nu$, flowing towards an equilateral-like shape in the vicinity of $\nu=0$. Intuitively, the close similarities with the scalar QsF sector are down to the fact that a $c_{\sigma}\ll 1$ (which is where we differ, given that $c^{\rm QsF}_{\sigma}= 1$) affects equally all of the wavefunctions $(k_1,k_2,k_3)$-dependence in the bispectrum calculation. The biggest impact of a small sound speed is then on the overall amplitude, not the shape profile.\\
Even before the actual calculation, and without the comparison with results from QsF, our findings are consistent with the following line of reasoning.\\
\noi In the presence of only light fields and their (self)interactions, the peak configuration of the bispectrum can be extracted from  identifying the type of interactions corresponding to each contribution: non-derivative interactions lead to a peak in the local configuration whilst increasing the number of derivative operators moves the profile towards equilateral or similar  (e.g. orthogonal) shapes. Derivative interactions may indeed introduce powers of $k_L/k_S$ that disfavour the squeezed configuration. There is something  to be added to this picture in the presence of massive states. Indeed, as pointed out elsewhere (see e.g.\cite{Assassi:2012zq}), the additional\footnote{Additional with respect to the late-time expansion in the massless case.} $(k\tau)^{3/2-\nu}$ late-time behaviour of massive wave-functions effectively adds a  $(k_L/k_S)^{f(3/2-\nu)}$ in front\footnote{The quantity $f$ here stands for a generic, typically positive, function.} of correlations functions and this has an effect on the overall shape that mimics the one due to the presence of derivative (self)interactions.\\

\noi A few comments are in order about the conventions used in plotting the shape-functions. First of all, we note that in all the plots a multiplying  factor of $(k_1 k_2 k_3)^2$ has been included in front of the expression for the bispectrum to conform to the standard definition of shape, see e.g.  \cite{Fergusson:2006pr}. Inspection of the RHS of Eq.(\ref{eee}) reveals the presence  of the momentum-conservation delta function, the quantity $B_{(\sigma)}$, and also the factor $\mathcal{A}^{\lambda_1 \lambda_2 \lambda_3}$ which encapsulates the effect of tensor helicities. In order to facilitate the comparison with the scalar sector of QsF, we will not be including the effect of $\mathcal{A}^{\lambda_1 \lambda_2 \lambda_3}$ in the shape-function plots. It is worth mentioning that for most combinations of the helicities this factor would, if included, have no qualitative influence on the profile, the exception being the squeezed configuration of the   combinations in the third row of Table~(\ref{table1}). We also stress that the expression in Eq.(\ref{helicities}) is different from the one characterising standard contributions (i.e. those not mediated by $\sigma$) to tensor non-Gaussianities and this fact may be used to distinguish the two scenarios.\\

\begin{figure}[h!]
\begin{center}
  \includegraphics[width=7.0cm]{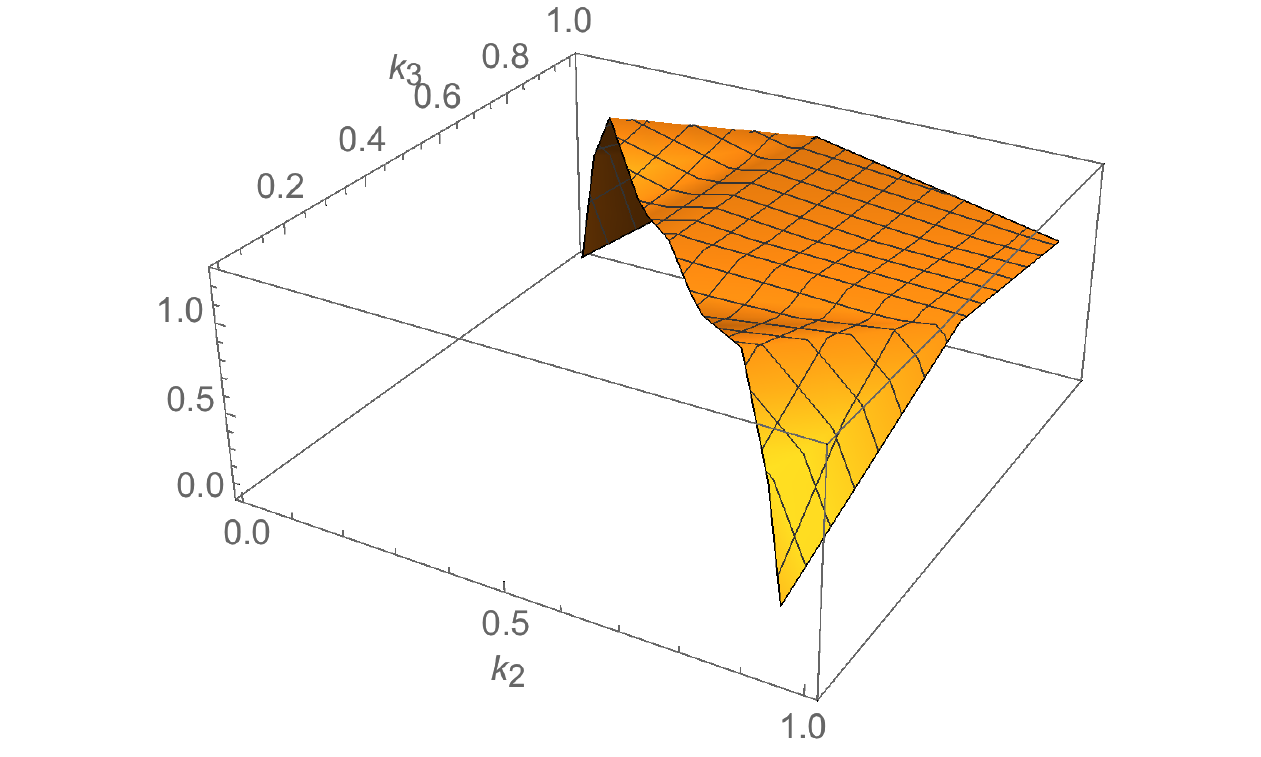}
  \caption
 {Plot of the numerically obtained shape-function $(k_1 k_2 k_3)^2 B_{\sigma}(k_1, k_2, k_3, c_{\sigma}, \nu)$ for $\nu=0,\, c_{\sigma}=1/10$ and where, without loss of generality, we have set  $k_1=1$. The value $\nu=0$  corresponds to $m= 3/2 \,H$. 
 }
 \label{fig11}
\end{center}
\end{figure}

\noi  We start with Fig.~(\ref{fig11}), where we plot the shape-function obtained numerically from the general expression in Eq.(\ref{fullb}) for the case $c_{\sigma}=1/10,\, \nu=0$. The profile that emerges is one with a plateau that quickly and smoothly decreases towards much smaller values in the squeezed region. The shape has been normalised according to its numerical value in the equilateral point of the $(k_2,k_3)$ plane. 

\begin{figure}[h!]
\begin{center}
  \includegraphics[width=5.5cm]{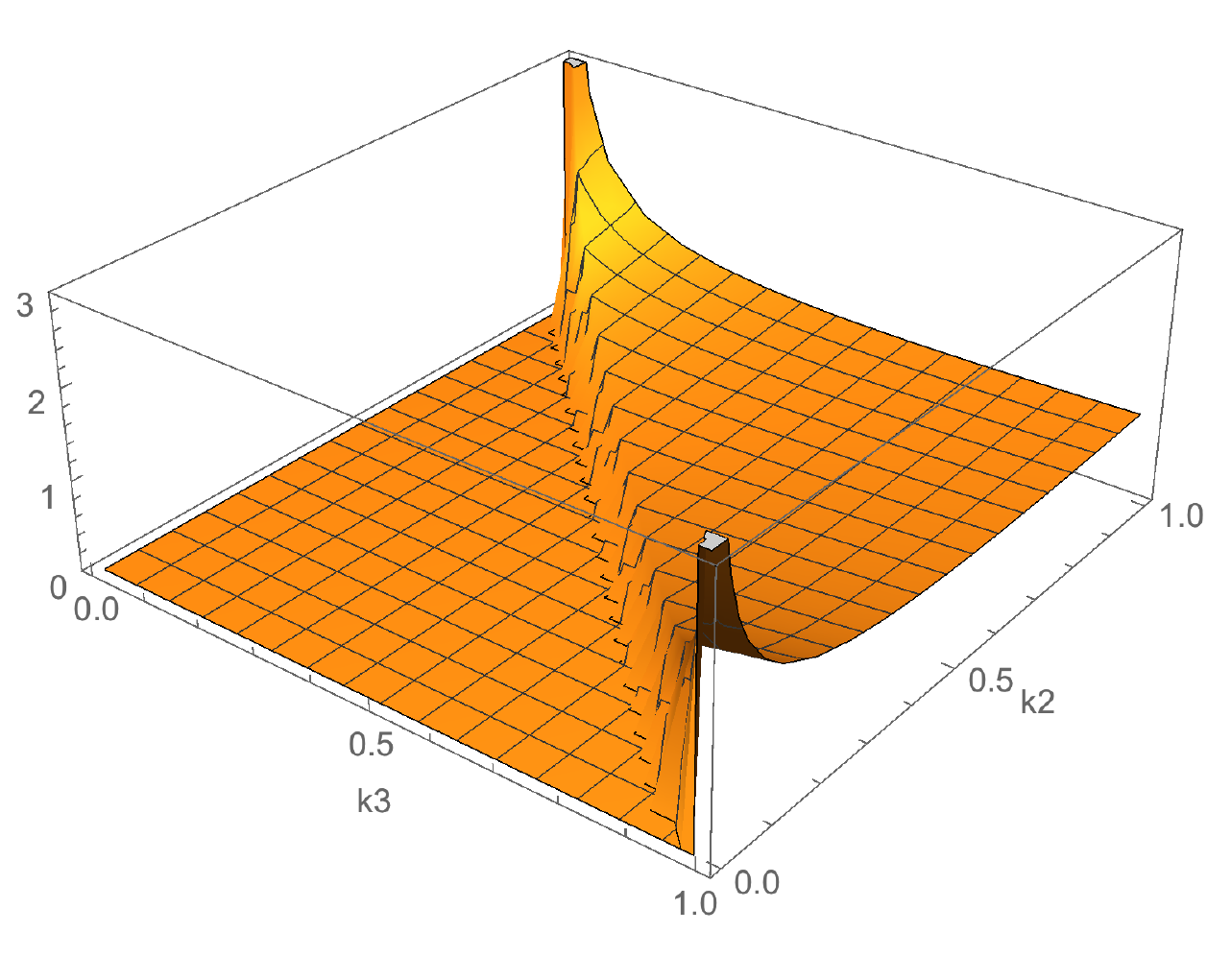}
  \caption
 {Plot corresponding to the shape-function $(k_1 k_2 k_3)^2 B_{\sigma}(k_1, k_2, k_3, c_{\sigma}, \nu) \Theta_{\rm Heav.}(k_2+k_3-1)$ for $\nu=1,\, c_{\sigma}=1/10$ and where, without loss of generality, we have set  $k_1=1$. The value $\nu=1$  corresponds to $m\simeq 1.12\,H$. The shape-function has been obtained by extrapolating to the full $(k_2,k_3)$ plane our squeezed limit findings. }
 \label{fig22}
\end{center}
\end{figure}

\noi A rather different shape-function corresponds to the $\nu=1$ case: we are flowing towards smaller values of the mass for the $\sigma$ field in parameter space so that the plot in Fig.~(\ref{fig22}) exhibits a clear peak in the squeezed limit. Irrespective of the results on the numerical integrals derived from  Eq.(\ref{fullb}), we can extract analytically the $\nu$-dependent behaviour of the momenta in the squeezed limit. It reads:
\bea
\label{squeezed}
B_{\sigma}(k_1,k_2,k_3,\nu)\Big|_{k_3\ll k_2\sim k_1}
\propto \left( \frac{1}{k_1^{7/2-\nu}\, k_2 \,k_3^{3/2+\nu} } \; + \text{perms.}\right) \; ,
\eea   
where we have omitted any dependence on the overall amplitude as well as on the function $s^{\text{sq}}(\nu,c_{\sigma})$.

\noi Strictly speaking, Eq.~(\ref{squeezed}) describes very accurately  the shape-function in two small regions centered on the $(0,1)$ and $(1,0)$ corners of the $(k_2,k_3)$ plane, but does not necessarily perform as well on the full plane. However, we have verified that, for the $\nu=1$ case, the shape obtained by extending the domain of Eq.~(\ref{squeezed}) is indeed a good approximation for the full numerical result and it is the former that we plot in Fig.~(\ref{fig22}).\\
\noi { It is worth mentioning at this stage the tensor bispectrum shape-function corresponding to other, specific, models with extra particle content. A case in point is the tensor sector of the axion-gauge field model in \cite{Dimastrogiovanni:2016fuu}. Non-Gaussianities were analyzed for the first time in \cite{Agrawal:2017awz}, followed by an even more detailed study \cite{Agrawal:2018mrg} by the same authors. The resulting bispectrum is peaked in the equilateral limit but, rather interestingly, it also displays a very distinct behaviour in other regions of the $(k_2,k_3)$ plane. This (class of) model(s) is of course different from the set-up under scrutiny here.  The most immediately identifiable difference at the level of observables is the  possibility of a peak in the squeezed configuration, as illustrated in Fig.~(\ref{fig22}). We intuitively explain this by the fact that the most intruiging parameter space region of the model analysed in \cite{Agrawal:2017awz,Agrawal:2018mrg}  corresponds to a larger effective mass for the extra fields than the one we are probing here for $\nu\gtrsim 1$. 
}

\subsection{On LISA response functions for tensor bispectra with intermediate shapes}
\label{lrf}

In the previous Sections, we learned that our scenario based on the EFT analysis of \cite{Bordin:2018pca}
can lead  to many possibilities for the momentum dependence corresponding to the shape of the tensor bispectrum.
 Bispectra can interpolate between squeezed and equilateral configurations, depending on the values of 
 parameters involved, and also other shapes are in principle possible.  It is interesting to start asking what are the  experimental prospects for distinguishing 
 the shape and chirality dependence  of the tensor bispectrum, in case of future detection of primordial gravitational waves from inflation. A systematic
 study of this problem by means of CMB polarisation is (as far as we are aware) still lacking, although  a detailed
 investigation   to analyse parity violating 
 tensor bispectra maximised in equilateral configurations  has been carried on by the Planck collaboration \cite{planckng},
 based on templates developed in \cite{Shiraishi:2011st}.

  In this section we start exploring how interferometers can be sensitive
 to the shape of tensor bispectrum, focussing on LISA \cite{Audley:2017drz}. As explained in the Introduction, there are
 examples of
 inflationary scenarios  with enhanced (w.r.t. the minimal case) amplitude of tensor power spectra and bispectra at interferometer
 scales. In case of direct couplings with additional spin two fields -- as in the EFT framework we are examining -- these scenarios
 can lead to new intermediate  shapes for  primordial tensor bispectra. The possibility of studying primordial tensor bispectra
 with LISA has been first  investigated in \cite{Bartolo:2018qqn}, where the amplitude of the three-point function for time delays of photons travelling along the interferometer
 arms is connected,  by means of the interferometer response function,  
with the amplitude and momentum dependence of 
 primordial tensor bispectra. The starting equation is 
 \begin{equation} \label{3ptref}
 \langle \Sigma^3(t) \rangle \,=\,\left(2 \pi L\right)^3\,\sum_{\lambda_1 \lambda_2 \lambda_3}
 \int_0^{\infty}
 \prod_{i=1,2,3}
 k_i d k_i\,{\cal B}_{\lambda_1 \lambda_2 \lambda_3}(\hat k_i^*, k_i)\,\,{\cal R}_{\lambda_1 \lambda_2 \lambda_3}(\hat k_i^*, k_i)
\,.
 \end{equation}
In the previous equation $\Sigma(t)$ is the time delay for a photon travelling along one of the
channels of LISA interferometer~\footnote{See \cite{Adams:2010vc} for 
 more details on the definition of LISA A, E, T channels,  corresponding to convenient combinations of interferometer signals that reduce the
 instrumental noise. In this Section, for simplicity we focus on the E channel.}. $L$ is the length of the interferometer arm.  
 ${\cal B}_{\lambda_1 \lambda_2 \lambda_3}(\hat k_i^*, k_i)$ is the tensor bispectrum
 including the helicity indexes (in our specific case, it reads ${\cal B}_{\lambda_1 \lambda_2 \lambda_3}\,
 =\,{\cal A}_{\lambda_1 \lambda_2 \lambda_3}  B_{(\sigma)}$ where the latter 
 quantities are defined in 
 Eqs.~\eqref{helicities}, \eqref{fullb}).  ${\cal R}_{\lambda_1 \lambda_2 \lambda_3}(\hat k_i^*, k_i)$ is the interferometer three-point response function. 
 The latter function depends on the size 
of three momenta, but also on the orientation of a  reference choice of three-vectors. See \cite{Bartolo:2018qqn} for full details on
Eq.~\eqref{3ptref} and on
 how to calculate the three-point response function.

Ref.~\cite{Bartolo:2018qqn} analysed in full detail the momentum dependence of the interferometer response function focussing  on the case of 
 primordial tensor bispectra maximised in the equilateral or in the squeezed configurations.  In this Section, we explore for the first time the detector response function to more general shapes of tensor bispectra. The shapes of tensor bispectra can be described
 in terms of a closed triangle in momentum shape, that for simplicity we consider in terms of   isosceles triangles formed by three 3-momenta ${\bf k}_i$, with angle $\alpha$ between the long sides. For example, 
 $\alpha=0$ corresponds to a squeezed isosceles triangle, $\alpha\,=\,\pi/3$ to an equilateral configuration, $\alpha\,=\,\pi$ to a
 folded  configuration.

\begin{figure}[h!]
    \hspace{40mm}
  \includegraphics[width=88mm]{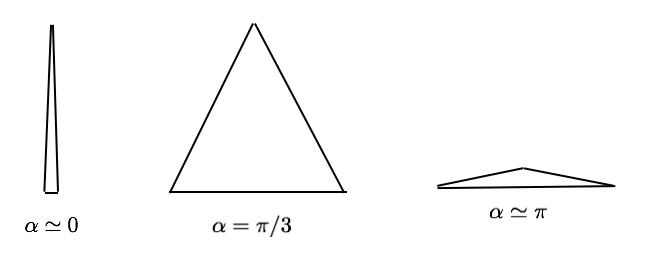}
\end{figure}

\begin{figure}[h!]
    \hspace{-15mm}
  \includegraphics[width=88mm]{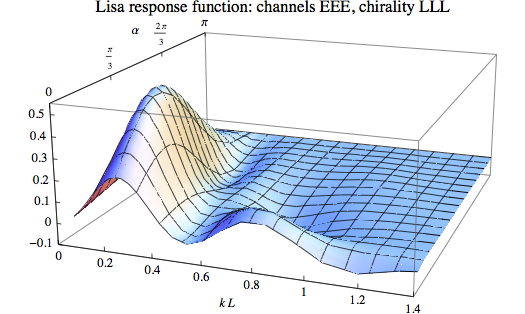}
  \hspace{.1mm}
  \includegraphics[width=88mm]{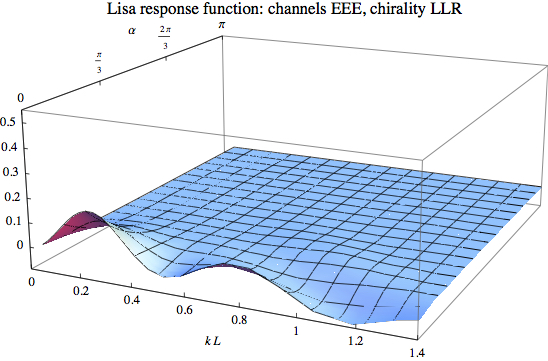}
  \caption
 {Amplitude of the LISA three-point response function ${\cal R}$ for signals travelling on the interferometer E channel. We provide the response functions for two choices of helicity combinations of the primordial tensor bispectrum, Left-Left-Left and Left-Left-Right. See the text for details. }
 \label{figLisa}
\end{figure}

Using the same methods and conventions of  \cite{Bartolo:2018qqn}, we represent in Figure \ref{figLisa} three dimensional plots for the  amplitude of the interferometer response functions, as function of the scale and of the angle $\alpha$.  The
  first plot  represents the amplitude of  the response functions relative to Left-Left-Left (LLL) helicity ${\cal R}_{LLL}$; 
 the other the amplitude for the  Left-Left-Right (LLR) component ${\cal R}_{LLR}$. The amplitude of
 the three point response functions for the remaining chirality indexes, or for other LISA channels, can be determined starting from our results by simple  analytical arguments \cite{Bartolo:2018qqn}.

 \bigskip
 
\bigskip

 Even a quick visual analysis of these plots allows us to derive some general physical consequences: we learn that in both cases, LLL and LLR, 
  the amplitude of the three-point response function smoothly depends on the shape of the triangle, while  strongly depends on the momentum scale one considers, with an oscillatory behaviour as a function of momenta. For the case of  LLL helicities, at a fixed momentum scale of order $k L\simeq 0.2$, the response function is maximal around  configurations near equilateral shapes (actually for $\alpha \simeq \pi/2$),  and smoothly decreases towards squeezed or folded configurations. The behavior is opposite for  LLR helicities, where squeezed configurations are enhanced, while the amplitude of the response function rapidly decreases towards equilateral and folded shapes. These considerations are interesting when applied to models of inflation with enhanced tensor fluctuations, and that priviledge some helicity components of the bispectrum with respect to others -- for example due to parity violating interactions of the inflaton with gauge fields. When accompanied by the interactions we consider with extra spin-2 fields, these scenarios can lead to a rich phenomenology for shape and chirality dependence of the tensor bispectrum. These can in principle be tested at interferometers exploiting the shape of momentum dependences of the instrument response function. A more detailed analysis of the capabilities of LISA to detect tensor non-Gaussianities of intermediate shapes -- including the computation of the corresponding signal-to-point ratio
 as done in \cite{Bartolo:2018qqn} for equilateral configurations -- is left for future work.
  
%
%

\smallskip

\section{Conclusions}
\label{sec:4}
The use of laser interferometers to detect gravitational waves has already proven spectacularly successful. Several detection events resulted in a wealth of new information: from bounds on the speed of propagation for gravitational
waves (and corresponding constraints on e.g. modified gravity models) to  signatures of the production of heavy elements, such as gold, as a result of neutron stars collision. Remarkably, besides celebrated CMB probes, also the newly acquired sensitivity to gravitational radiation at interferometer scales has the potential to inform us  about one of the earliest epochs in the universe history: inflation. The universe make-up during this early acceleration phase may indeed be imprinted in the correlations functions of tensor fluctuations of the metric field.\\
\indent The simplest inflationary scenarios posit a PGW spectrum which is a slightly decreasing function of the wavenumber $k$ and deliver small tensor higher-point statistics.
Extrapolating from known upper bounds on the signal at CMB scales, one may conclude that the standard single-field paradigm would evade constraints from both LIGO and LISA. Intriguingly, a much more dramatic outcome is predicted by entire classes of inflationary models. What these have in common is the existence of a rich particle content during inflation. As we have seen, it is rather useful in this context to classify particles according to their spin: the corresponding unitarity constraints will reflect directly on their mass range and, in turn, this will affect their impact on cosmological  observables, including the gravitational waves spectrum and  higher-order correlations functions, i.e. tensor non-Gaussianities. Higher-spin fields are most affected, so much so that non-minimal coupling with the  inflaton sector is typically necessary to enhance their impact on the tensor sector all the way to e.g. a LISA detection. A PGW detection (or lack theoreof)  at intereferometer scales then inevitably tells us about the mass, the spin and also the coupling of the inflationary particle zoo. \\
\indent In this work, we have adopted the EFT approach of \cite{Bordin:2018pca} to investigate in all generality the imprints on  primordial tensor non-Gaussianities of an extra spin-2 field with sizable coupling to the inflaton field.  We find that, as expected, a small sound speed for the extra field, $\sigma$, delivers a large signal. The corresponding bispectrum shape-function interpolates, as one decreases the mass $m_{\sigma}$, from the equilateral to the local profile, in complete analogy with non-Gaussianities in the scalar sector of quasi-single-field \cite{Chen:2009zp} inflation. \\
\indent Given that a blue tensor spectrum is allowed within the EFT set-up, we also provide an estimate of the response function of the LISA interferometer in momentum configurations that interpolate between the local, the equilateral, and the flattened. We do so for two combinations of the helicities, LLL and LLR, finding that the two respond differently to different momenta configurations. These are promising results that call for a more detailed characterization of the capabilities of LISA to constrain tensor non-Gaussianities of intermediate shapes.

\section*{Acknowledgements}
ED, MRF, and especially GT, would like to thank the LISA Cosmology Working Group for many discussions on related subjects. ED is supported in part by DOE grant DE-SC0009946. MRF and DW acknowledge support from STFC grant ST/N000668/1. GT is partially supported by the STFC grant ST/P00055X/1.

\begin{appendix}

\section{Bispectrum computation} 
\label{AppendixA}

\noindent The equation of motion for $\sigma_{k}$ to leading order in slow-roll reads 
\begin{equation}
\sigma_{k}^{''}-\frac{2}{\tau}\sigma_{k}^{'}+\left(c_{\sigma}^{2}k^{2}+\frac{m^{2}}{H^{2}\tau^{2}}\right)\sigma_{k}=0\,,
\end{equation}
whose solution is given in Eq.~(\ref{solsigma}).\\

\noindent The diagram in Fig.~\ref{fig1} is computed using the in-in formula (see e.g. \cite{Weinberg:2005vy,Adshead:2009cb})
\begin{equation}\label{inin}
\left\langle \gamma^3 \right\rangle=\left\langle 0\,\Big| \left[\bar{T}\,\exp\left(i\int_{t_{0}}^{t}dt^{'}H_{I}(t^{'})\right)\right]  \gamma^3_{I}\left[T\,\exp\left(-i\int_{t_{0}}^{t}dt^{''}H_{I}(t^{''})\right)\right] \Big|\,0 \right\rangle\,,
\end{equation}
where the subscript $I$ stands for fields in the interaction picture, $\bar{T}$ and $T$ are, respectively, anti-time ordering and time-ordering operators and the interaction Hamiltonian is given by
\begin{equation}
\mathcal{H}=\mathcal{H}_{2}+\mathcal{H}_{3}\,,\quad\quad \mathcal{H}_{2}=-\frac{\rho\,M_{\text{Pl}}}{2}\,a^3\sigma^{ij}\dot{\gamma}_{ij} \,,\quad\quad\mathcal{H}_{3}=\mu\, a^3\,\sigma_{ij}^{3}\,.
\end{equation}
The Fourier expansion of the spin-2 field fluctuations was given in Eq.~(\ref{fexp1}), where 
\begin{eqnarray}\label{fexp}
&&\gamma^{\lambda}_{\textbf{k}}(\tau)=a_{\textbf{k}}^{\lambda}\gamma_{k}^{\lambda}(\tau)+a_{-\textbf{k}}^{\lambda\dagger}\gamma_{k}^{\lambda *}(\tau)\,,\\
&&\sigma^{\lambda}_{\textbf{k}}(\tau)=b_{\textbf{k}}^{\lambda}\sigma_{k}^{\lambda}(\tau)+b_{-\textbf{k}}^{\lambda\dagger}\sigma_{k}^{\lambda *}(\tau)\,, 
\end{eqnarray}
with creation and annihilation operators obeying the commutation relations
\begin{equation}
\left[a^{\lambda_{1}}_{\textbf{k}_{1}},a^{\lambda_{2}\,\dagger}_{-\textbf{k}_{2}}\right]=(2\pi)^3\delta_{\lambda_{1}\lambda_{2}}\delta^{(3)}(\textbf{k}_{1}+\textbf{k}_{2})\,\quad\quad \left[b^{\lambda_{1}}_{\textbf{k}_{1}},b^{\lambda_{2}\,\dagger}_{-\textbf{k}_{2}}\right]=(2\pi)^3\delta_{\lambda_{1}\lambda_{2}}\delta^{(3)}(\textbf{k}_{1}+\textbf{k}_{2})\,.
\end{equation}
In order to quickly illustrate the calculations, we report below the 
schematic form of the bispectrum
\begin{equation}
\langle \gamma_{\textbf{k}_{1}}^{\lambda_{1}} \gamma_{\textbf{k}_{2}}^{\lambda_{2}} \gamma_{\textbf{k}_{3}}^{\lambda_{3}} \rangle\sim\gamma_{\textbf{k}_{1}}^{\lambda_{1}} \gamma_{\textbf{k}_{2}}^{\lambda_{2}} \gamma_{\textbf{k}_{3}}^{\lambda_{3}}\,\,\Pi_{n=1,2,3}\int d^{4}x_{n}\,a^{3}(\tau_{n}) \dot{\gamma}_{ij}(x_{n})\,\sigma^{ij}(x_{n})\int d^{4}y\,\,a^{3}(\tau_{y}) \sigma_{ij}^{3}(y)\,.\nonumber\\
\end{equation}
Each contraction of the $\gamma$ fields with one another yields delta functions in momentum and helicity (and likewise for the $\sigma$ fields),  reducing the initial nine polarization tensors down to three, hence the coefficient $\epsilon^{\lambda_{1}}_{\alpha\beta}(-\hat{k}_{1})\epsilon^{\lambda_{2}}_{\beta\gamma}(-\hat{k}_{2})\epsilon^{\lambda_{3}}_{\gamma\alpha}(-\hat{k}_{3})$ in Eq.~(\ref{eee}). In deriving the latter we made use of the following relations: 
\begin{equation}
\epsilon^{L/R}_{ij}(\hat{k})\cdot\epsilon^{R/L}_{ij}(\hat{k})=1\,,\quad\quad\epsilon^{L/R}_{ij}(\hat{k})\cdot\epsilon^{L/R}_{ij}(\hat{k})=0\,,
\end{equation}
where $\epsilon^{R/L*}_{ij}(\hat{k})=\epsilon^{R/L}_{ij}(-\hat{k})=\epsilon^{L/R}_{ij}(\hat{k})$. One can rewrite the overall coefficient as follows
\begin{eqnarray}\label{hel}
\epsilon^{\lambda_{1}}_{\alpha\beta}(\hat{k}_{1})\epsilon^{\lambda_{2}}_{\beta\gamma}(\hat{k}_{2})\epsilon^{\lambda_{3}}_{\gamma\alpha}(\hat{k}_{3})&=&-\frac{1}{64\,r_{2}^{2}r_{3}^{2}}\left[2r_{2}+\left(1+r_{2}^{2}-r_{3}^{2}\right)\lambda_{1}\lambda_{2}\right]\left[-2r_{3}+\left(-1+r_{2}^{2}-r_{3}^{2}\right)\lambda_{1}\lambda_{3}\right]\nonumber\\&&\quad\quad\times\left[2r_{2}r_{3}+\left(-1+r_{2}^{2}+r_{3}^{2}\right)\lambda_{2}\lambda_{3}\right]
\end{eqnarray}
where $r_{2}\equiv k_{2}/k_{1}$ and $r_{3}\equiv k_{3}/k_{1}$. Eq.~(\ref{hel}) has been obtained by choosing a reference triangle for the wave vectors that sits on the $(x,y)$ plane: 
\begin{equation}
\epsilon^{R/L}_{ij}\left(\theta=\frac{\pi}{2},\varphi\right)=\frac{1}{2}\left[ {\begin{array}{ccc}
   -\sin^2\varphi\,\,\, & \cos\varphi \sin\varphi\,\,\, & \pm i \sin\varphi\\
   \cos\varphi \sin\varphi\,\,\,&  -\cos^2\varphi\,\,\,& \mp i \cos\varphi\\
     \pm i \sin\varphi\,\,\,& \mp i \cos\varphi\,\,\,& 1\\
  \end{array} } \right]\,.
\end{equation}
One can set $\varphi_{1}$ to zero, hence $\varphi_{2}$ and $\varphi_{3}$ are the angles between, respectively, $\textbf{k}_{2}$ and $\textbf{k}_{1}$ and $\textbf{k}_{3}$ and $\textbf{k}_{1}$.\\

\noindent The expression for the bispectrum is given by the sum of three contributions (see also \cite{Chen:2009zp})
\begin{equation}
B_{(\sigma)}(k_{1},k_{2},k_{3})=B^{(A)}(k_{1},k_{2},k_{3})+B^{(B)}(k_{1},k_{2},k_{3})+B^{(C)}(k_{1},k_{2},k_{3})
\end{equation}
where 
\begin{eqnarray}\label{a123general}
B^{(A)}&=&12\cdot\pi^3\left(\frac{\mu}{H}\right)\left(\frac{\rho}{M_{\text{Pl}}}\right)^{3}\frac{1}{k_{1}k_{2}k_{3}}\int_{-\infty}^{0} d\tau_{1}\int_{-\infty}^{\tau_{1}} d\tau_{2}\int_{-\infty}^{\tau_{2}} d\tau_{3}\int_{-\infty}^{\tau_{3}} d\tau_{4} \sqrt{\frac{\tau_{2}}{\tau_{1}\tau_{3}\tau_{4}}} \\&&\times \text{Im}\left[e^{i k_{2}\tau_{3}}\mathcal{H}^{(1)}_{\nu}[-c_{\sigma}k_{2}\tau_{2}]\mathcal{H}^{(2)}_{\nu}[-c_{\sigma}k_{2}\tau_{3}]\right]\text{Im}\left[\mathcal{H}^{(1)}_{\nu}[-c_{\sigma}k_{1}\tau_{1}]\mathcal{H}^{(2)}_{\nu}[-c_{\sigma}k_{1}\tau_{2}]\right]\nonumber\\&&\times\left[e^{-i k_{3}\tau_{4}}\mathcal{H}^{(1)}_{\nu}[-c_{\sigma}k_{3}\tau_{4}]\mathcal{H}^{(2)}_{\nu}[-c_{\sigma}k_{3}\tau_{2}]\right]\sin[-k_{1}\tau_{1}]+\text{5\,perms} \,,\nonumber
\end{eqnarray}
\begin{eqnarray}
\label{b123general}
B^{(B)}&=&12\cdot\pi^3\left(\frac{\mu}{H}\right)\left(\frac{\rho}{M_{\text{Pl}}}\right)^{3}\frac{1}{k_{1}k_{2}k_{3}}\int_{-\infty}^{0} d\tau_{1}\int_{-\infty}^{\tau_{1}} d\tau_{2}\int_{-\infty}^{\tau_{2}} d\tau_{3}\int_{-\infty}^{\tau_{3}} d\tau_{4} \sqrt{\frac{\tau_{3}}{\tau_{1}\tau_{2}\tau_{4}}}
\\&&\times \text{Im}\left[\mathcal{H}^{(1)}_{\nu}[-c_{\sigma}k_{1}\tau_{3}]\mathcal{H}^{(1)}_{\nu}[-c_{\sigma}k_{2}\tau_{3}]\mathcal{H}^{(2)}_{\nu}[-c_{\sigma}k_{1}\tau_{1}]\mathcal{H}^{(2)}_{\nu}[-c_{\sigma}k_{2}\tau_{2}]\right] 
\nonumber\\&&\times \text{Im}\left[e^{i k_{3}\tau_{4}}\mathcal{H}^{(1)}_{\nu}[-c_{\sigma}k_{3}\tau_{3}]\mathcal{H}^{(2)}_{\nu}[-c_{\sigma}k_{3}\tau_{4}]\right] \sin[-k_{1}\tau_{1}]\sin[-k_{2}\tau_{2}]   +\text{5\,perms}\,,\nonumber
\end{eqnarray}
\begin{eqnarray}
\label{c123general}
B^{(C)}&=&12\cdot\pi^3\left(\frac{\mu}{H}\right)\left(\frac{\rho}{M_{\text{Pl}}}\right)^{3}\frac{1}{k_{1}k_{2}k_{3}}\int_{-\infty}^{0} d\tau_{1}\int_{-\infty}^{\tau_{1}} d\tau_{2}\int_{-\infty}^{\tau_{2}} d\tau_{3}\int_{-\infty}^{\tau_{3}} d\tau_{4} \sqrt{\frac{\tau_{4}}{\tau_{1}\tau_{2}\tau_{3}}}\nonumber
\\&&\times\text{Im}\Big[\mathcal{H}^{(1)}_{\nu}[-c_{\sigma}k_{1}\tau_{4}]\mathcal{H}^{(1)}_{\nu}[-c_{\sigma}k_{2}\tau_{4}]\mathcal{H}^{(1)}_{\nu}[-c_{\sigma}k_{3}\tau_{4}]\mathcal{H}^{(2)}_{\nu}[-c_{\sigma}k_{1}\tau_{1}]\\&&\times\mathcal{H}^{(2)}_{\nu}[-c_{\sigma}k_{2}\tau_{2}]\mathcal{H}^{(2)}_{\nu}[-c_{\sigma}k_{3}\tau_{3}] \Big]  \sin[-k_{1}\tau_{1}]\sin[-k_{2}\tau_{2}]\sin[-k_{3}\tau_{3}]+\text{5\,perms}\,.\nonumber
\end{eqnarray}
From here on it is convenient to introduce the variables: $x_{i}\equiv k_{1}\tau_{i}$. The complete expression of the bispectrum then becomes (\ref{fullb}), with $\mathcal{F}_{\nu c_{\sigma}}=\mathcal{F}^{A}_{\nu c_{\sigma}}+\mathcal{F}^{B}_{\nu c_{\sigma}}+\mathcal{F}^{C}_{\nu c_{\sigma}}$,
where
\begin{eqnarray}\label{A123general}
\mathcal{F}^{A}_{\nu c_{\sigma}}&\equiv&\int\Pi_{i}dx_{i}\, \sqrt{\frac{x_{2}}{x_{1}x_{3}x_{4}}}\sin[-x_{1}]\text{Im}\left[e^{-i(k_{3}/k_{1})x_{4}}\mathcal{H}^{(1)}_{\nu}[-c_{\sigma}(k_{3}/k_{1})x_{4}]\mathcal{H}^{(2)}_{\nu}[-c_{\sigma}(k_{3}/k_{1})x_{2}]\right]\nonumber\\&&\text{Im}\left[e^{i(k_{2}/k_{1})x_{3}}\mathcal{H}^{(1)}_{\nu}[-c_{\sigma}(k_{2}/k_{1})x_{2}]\mathcal{H}^{(2)}_{\nu}[-c_{\sigma}(k_{2}/k_{1})x_{3}]\right]\text{Im}\left[\mathcal{H}^{(1)}_{\nu}[-c_{\sigma}x_{1}]\mathcal{H}^{(2)}_{\nu}[-c_{\sigma}x_{2}]\right]
\,,\nonumber\\\quad\\
%
\label{B123general}
\mathcal{F}^{B}_{\nu c_{\sigma}}&\equiv& -\int\Pi_{i}dx_{i}\,\sqrt{\frac{x_{3}}{x_{1}x_{2}x_{4}}}\text{Im}\left[\mathcal{H}^{(1)}_{\nu}[-c_{\sigma}x_{3}]\mathcal{H}^{(1)}_{\nu}[-c_{\sigma}(k_{2}/k_{1})x_{3}]\mathcal{H}^{(2)}_{\nu}[-c_{\sigma}x_{1}]\mathcal{H}^{(2)}_{\nu}[-c_{\sigma}(k_{2}/k_{1})x_{2}]\right]\nonumber\\&&\text{Im}\left[e^{i(k_{3}/k_{1})x_{4}}\mathcal{H}^{(1)}_{\nu}[-c_{\sigma}(k_{3}/k_{1})x_{3}]\mathcal{H}^{(2)}_{\nu}[-c_{\sigma}(k_{3}/k_{1})x_{4}]\right]\sin[-x_{1}]\sin[-(k_{2}/k_{1})x_{2}],\textcolor{white}{niiiiin}\\
%
\label{C123general}
\mathcal{F}^{C}_{\nu c_{\sigma}}&\equiv &\int\Pi_{i}dx_{i}\,\sqrt{\frac{x_{4}}{x_{1}x_{2}x_{3}}}\sin[-x_{1}]\sin[-(k_{2}/k_{1})x_{2}]\sin[-(k_{3}/k_{1})x_{2}] \nonumber \\&&\text{Im}\Big[\mathcal{H}^{(1)}_{\nu}[-c_{\sigma}x_{4}]\mathcal{H}^{(1)}_{\nu}[-c_{\sigma}(k_{2}/k_{1})x_{4}]\mathcal{H}^{(1)}_{\nu}[-c_{\sigma}(k_{3}/k_{1})x_{4}]\mathcal{H}^{(2)}_{\nu}[-c_{\sigma}x_{1}]\\&&\quad\quad\times\mathcal{H}^{(2)}_{\nu}[-c_{\sigma}(k_{2}/k_{1})x_{2}]\mathcal{H}^{(2)}_{\nu}[-c_{\sigma}(k_{3}/k_{1})x_{3}]\Big]\,,\nonumber
\end{eqnarray}
where $\int\Pi_{i}dx_{i}\equiv\int_{-\infty}^{0}dx_{1}\int_{-\infty}^{x_{1}}dx_{2}\int_{-\infty}^{x_{2}}dx_{3}\int_{-\infty}^{x_{3}}dx_{4}$. In the squeezed and equilateral configurations, our results simplify and the momentum dependence can be factored out to arrive respectively at Eqs.~(\ref{sq1}) and Eqs.(\ref{seq1}). Eq.~(\ref{seq1}) is an exact result. The squeezed limit expression, (\ref{sq1}), is instead obtained by approximating some of the Hankel functions for small arguments. Consider for instance the contribution in Eq.~(\ref{A123general}) in the limit $k_{1}\sim k_{2}\gg k_{3}$. The second Hankel in the first line of (\ref{A123general}) can be approximated for small values of its argument ($c_{\sigma}(k_{3}/k_{1})|x_{2}|\ll 1$):
\begin{equation}
\mathcal{H}^{(2)}_{\nu}[-c_{\sigma}(k_{3}/k_{1})x_{2}]\simeq i\,\frac{2^{\nu}\,\Gamma[\nu]}{\pi\,c_{\sigma}^{\nu}}\left(\frac{k_{1}}{k_{3}}\right)^{\nu}(-x_{2})^{-\nu}\,.
\end{equation}
This expansion is justified by the fact that values $c_{\sigma}(k_{3}/k_{1})|x_{2}|\gtrsim 1$, i.e. $c_{\sigma}|x_{2}|\gg 1$, would cause some of the Hankel in the second line, specifically those of argument $c_{\sigma}(k_{2}/k_{1})x_{2}\simeq c_{\sigma}x_{2}$, to undergo fast oscillations, suppressing the value of the integral. Following similar reasonings, it is also possible to isolate one of the chain integrals (as the result in (\ref{sq1}) shows, where we set $y_{4}\equiv(k_{3}/k_{1})x_{4}$). We verified that the leading contribution in the squeezed limit arises from the permutations spelled out in (\ref{A123general}) and (\ref{B123general}), whereas the (two, independent) remaining permutations, as well as the C-type terms as in (\ref{C123general}), are subleading.

\end{appendix}

\end{document}